\documentclass{aa} 
\usepackage{txfonts}

\usepackage{graphicx,epsf}
\usepackage{amsmath,graphicx,adjustbox,setspace, sidecap, float}
\usepackage[toc,page]{appendix}
\usepackage{gensymb}
\usepackage{hyperref}
\usepackage{fancyhdr}
\usepackage{natbib}
\usepackage{caption,subcaption}
\usepackage{xcolor}
\usepackage{physics}
\usepackage{ulem}
\usepackage{multirow}

\newcommand{\sect}[1]{Sect.~\ref{sec:#1}}

\newcommand{\Sects}[2]{Sections\ \ref{sec:#1} and \ref{sec:#2}}

\newcommand{\fg}[1]{Fig.~\ref{fig:#1}}
\newcommand{\Fg}[1]{Figure~\ref{fig:#1}}

\newcommand{\eq}[1]{Eq.~(\ref{eq:#1})\xspace}
\newcommand{\Eq}[1]{Equation~(\ref{eq:#1})\xspace}

\begin{document}
\title{Spatial distribution of crystalline silicates in protoplanetary disks: How to interpret mid-infrared observations}

   \author{Hyerin Jang\inst{1}
          \and
          L.B.F.M. Waters\inst{1,2}
          \and
          I. Kamp\inst{3}
          \and
          C. P. Dullemond\inst{4}}

   \institute{Department of Astrophysics/IMAPP, Radboud University, PO Box 9010, 6500 GL Nijmegen, The Netherlands
   \newline\email{hyerin.jang@astro.ru.nl}
   \and
   SRON Netherlands Institute for Space Research, Niels Bohrweg 4, NL-2333 CA Leiden, the Netherlands
   \and 
   Kapteyn Astronomical Institute, University of Groningen, PO BOX 800, 9700 AV Groningen, The Netherlands
   \and
   Institute for Theoretical Astrophysics, Center for Astronomie (ZAH), Heidelberg University, Albert Ueberle Str. 2, 69120 Heidelberg, Germany}

   \date{\today}
    
\abstract
{\textit{Context.}  Crystalline silicates are an important tracer of the evolution of dust, the main building block of planet formation. In an inner protoplanetary disk, amorphous silicates are annealed because of the high temperatures that prevail there. These crystalline silicates are radially and vertically distributed by disk turbulence and/or radial transport. Mid-infrared spectrographs are sensitive to the presence and temperature of micron-sized silicates, and the dust temperature can be used to infer their spatial distribution. 

\textit{Aims.} We aim to model the spatial distribution of crystalline silicate dust in protoplanetary disks taking into account thermal annealing of silicate dust and radial transport of dust in the midplane. Using the resulting spatial distribution of crystalline and amorphous silicates, we calculate mid-IR spectra to study the effect on dust features and to compare these to observations.

\textit{Methods.} We modeled a Class II T-Tauri protoplanetary disk and defined the region where crystallization happens by thermal annealing processes from the comparison between crystallization and residence timescales ($\tau_{\text{cryst}} < \tau_{\text{res}}$). Radial mixing and drift were compared to find a vertically well mixed region ($\tau_{\text{ver}} < \tau_{\text{drift}}$). We used the DISKLAB code to model the radial transport in the midplane and obtained the spatial distribution of the crystalline silicates for different grain sizes. We used MCMax, a radiative transfer code, to model the mid-infrared spectrum.

\textit{Results.} In our modeled T-Tauri disk, different grain sizes get crystallized in different radial and vertical ranges within 0.2 au. Small dust gets vertically mixed up efficiently, so crystallized small dust in the disk surface is well mixed with the midplane. Inwards of 0.075 au, all grains are fully crystalline irrespective of their size. We also find that the crystallized dust is distributed out to a few au by radial transport, smaller grains more so than larger ones. Our fiducial model shows different contributions of the inner and outer disk to the dust spectral features. The $10~ \mu$m forsterite feature has $\sim 30 \%$ contribution from the innermost disk (0.07-0.09 au) and $<1 \%$ from the disk beyond 10 au while the $33~ \mu$m feature has $\sim 10 \%$ contribution from both innermost and outer disks. We also find that feature strengths change when varying the spatial distribution of crystalline dust. Our modeled spectra qualitatively agree with observations from the Spitzer Space Telescope, but the modeled 10 $\mu$m feature is strongly dominated by crystalline dust, unlike observations. Models with reduced crystallinity and depletion of small crystalline dust within 0.2 au show a better match with observations.

\textit{Conclusions.} Mid-infrared observations of the disk surface represent the radial distribution of small dust grains in the midplane and provide us with abundances of crystalline and amorphous dust, size distribution, and chemical composition in the inner disk. The inner and outer disks contribute more to shorter and longer wavelength features, respectively. In addition to the crystallization and dynamical processes, amorphization, sublimation of silicates, and dust evolution have to be taken into account to match observations, especially at $\lambda = 10 \mu$m, where the inner disk mostly contributes. This study could interpret spectra of planet forming disks taken with the Mid-Infrared Instrument (MIRI) on board the James Webb Space Telescope.

}
\keywords{methods: analytical -- protoplanetary disks -- infrared: planetary systems}
\maketitle

\section{Introduction}
\label{sec:intro}
\paragraph{}
Dust in protoplanetary disks is the main building block for the formation of terrestrial planets or cores of giant planets \citep{Raymond_Morbidelli2022}. Silicates are the most abundant dust component in interstellar clouds and planet-forming disks \citep{Dorschner_etal1995, Colangeli_etal2003,Henning2010}, along with carbonaceous dust. Crystalline silicates have been found in solar system comets \citep{Hanner_etal1994,Wooden2002}, interplanetary dust particles \citep{MacKinnon_Rietmeijer1987,Bradley_etal1992}, planet-forming disks, and the outflows of evolved stars \citep{Waters1996}. Because crystallization requires distinct conditions, crystalline silicates can be a tracer for the evolution of dust in a planet-forming disk, especially, for terrestrial planets in the inner disk. 

Significant amounts of these crystalline silicates have been observed in the infrared spectra of protoplanetary disks \citep{Bouwman_etal2001,vanBoekel_etal2004,vanBoekel_etal2005, KesslerSilacci_etal2005,Apai_etal2005, KesslerSilacci_etal2006, Juhasz_etal2010,Sturm_etal2013} while amorphous silicates are observed in diffuse interstellar medium \citep{Kemper_etal2004,Chiar_Tielens2006}. The abundance of warm and small crystalline silicates (summing up the different minerals that are commonly detected, e.g. forsterite Mg$_2$SiO$_4$ and enstatite MgSiO$_3$) varies greatly between disks, from upper limits of a few percent to ~30 \% or more in spatially unresolved full disk spectra \citep{Bouwman_etal2001,vanBoekel_etal2005,Apai_etal2005,Juhasz_etal2010}. Spatially resolved interferometric observations showed that the inner disks (1-2 au) of Herbig Ae stars have 40-95 \% crystallinity \citep{vanBoekel_etal2004,Menu_etal2015,Varga_etal2018,Varga_etal2019,Varga_etal2021}. On the other hand, amorphous silicates have been observed from the diffuse interstellar medium. By using infrared spectroscopy, \cite{Kemper_etal2004} reported that the 10 $\mu$m silicate absorption band toward the galactic center shows mostly amorphous pyroxene and olivine and no crystalline silicates. \cite{Chiar_Tielens2006} found that a mixture of olivine and pyroxene matches the spectrum of the sightline towards the Wolf-Rayet star WR98a. Based on these observations, the current hypothesis is that amorphous silicates in the interstellar medium are crystallized in the protoplanetary disk. 

Possible mechanisms for the formation of crystalline silicates are thermal annealing at high temperature or high temperature gas phase condensation of silicates. These conditions are likely to occur in the inner regions of protoplanetary disks due to direct radiation or outburst \citep{Abraham_etal2009} from the central star. Additionally, lightning \citep{Desch_Cuzzi2000} anywhere in the disk may locally create conditions favorable for the formation of crystalline silicates. On the other hand, crystallization of silicates is also possible to occur at low temperatures by a surface reaction \citep{Kaito_etal2007,Tanaka_etal2010} or under high pressure of water vapor, a significant reactive gas \citep{Yamamoto_Tachibana2018}. \cite{Bouwman_etal2003} also have detected crystalline silicates in the outer disk at $\sim 100$ K in HD 100546. However, we focus on thermal annealing process at high temperatures for crystallization in this paper.

Thermal annealing of amorphous silicates has been studied in laboratory experiments; crystallization depends on the given temperature and the duration of heating \citep{Hallenbeck_etal1998, Fabian_etal2000,Jager_etal2003, Thompson_etal2019} but also on the chemical composition of the silicate. For instance, \cite{Fabian_etal2000} showed that MgSiO$_3$ smoke gets crystallized within 3-5 minutes at 1121 K and longer than 50 hours at 1000 K. The higher the temperature, the shorter the duration for complete crystallization. Thus, the inner disk, where the temperature is high, is the favorable region for crystallization. 

A protoplanetary disk has both radial and vertical temperature profiles; thus, dust experiences different temperatures at different radial distances from the central star and vertical heights from the midplane. Dust dynamically travels within the disk due to drift by gas drag, accretion towards the central star, diffusion along with the gas, and disk turbulence, and experiences different temperatures. Thus, the abundance of crystalline silicates depends on the local conditions as well as on the transport of dust. In addition, these crystalline silicates are also observed beyond a few au where the disk temperature is not high enough for crystallization \citep{Bouwman_etal2001, Watson_etal2009, deVries_etal2012, Sturm_etal2013}. One possibility would be the dynamics of dust in the disk. Crystallized dust in the innermost disk gets radially transported to the cold region along with diffusing gas \citep{Gail2001,Wehrstedt_Gail2002,Desch2007,Desch_etal2017}. Dust winds could also transport small dust from the inner disk to the outer disk \citep{Miyake_etal2016,Giacalone_etal2019,Franz_etal2020,Rodenkirch_Dullemond2022}.

Crystalline silicates are easily detected at mid-infrared wavelengths because of the strong vibrational resonances that occur at these wavelengths. These resonances are sensitive to the chemical composition and temperature so can be used to probe the disk mineralogy. Because the resonances cover a wide wavelength range, information about the spatial distribution of crystalline silicates can be inferred from spatially unresolved data. This can be useful in combination with spatially resolved interferometric data, that are much more difficult to obtain due to observational detection limits. Especially, the Spitzer Space Telescope and more recently the James Webb Space Telescope (JWST) observed many protoplanetary disks in the mid-infrared wavelength.

However, at these wavelengths, we only observe the disk surface layers due to the high optical depth that is dominated by small, (sub) micron-sized dust grains. The dust at the disk surface represents a minor fraction of the solid reservoir and only the smallest grains are mixed up to the observable disk region. Therefore, the question arises if the grain composition and crystallinity of small grains at the disk surface are representative of those in the midplane. Terrestrial planets are formed in the midplane and the inner disk. Thus, it is important to study the spatial distribution of crystalline silicates in the midplane of the inner disk in order to better understand the dust as a tracer for planet formation. If we analytically understand how the spatial distribution of dust in the disk surface represents the midplane, we can infer from JWST data implication for planet formation. Previous studies \citep{Gail2001,Wehrstedt_Gail2002} focused on modeling the spatial distribution of crystalline silicate in planet forming disks. \cite{Maaskant_etal2015} investigated the impact of grain growth, disk gaps, radial mixing, and optical depth on the infrared features of forsterite using radiative transfer modeling. However, that study used a parametrized approach for crystallinity in disk zones and did not take into account the spatial distribution of different sizes of forsterite grains. In this study, for the first time, we model the spatial distribution of crystalline silicates and resulting mid-IR spectra to guide the interpretation of spatially unresolved observations at these wavelengths.

This paper is organized as follows: the dynamical model and radiative transfer model are introduced in \sect{method}. The spatial distribution of crystalline silicates and their modeled spectra with various parameters are shown in \sect{dist2spec}. In \sect{qual_comparison}, we compare our results with previous observations and suggest possible mechanisms to explain the observed silicate features. We also discuss our findings with different analytical and observational studies, values of parameters in the models, and potential further study in \sect{discussion}. Conclusions are summarized in \sect{conclusion}.

\section{Method}
\label{sec:method}
\paragraph{}
We first introduce our T-Tauri disk model in Sect. \ref{sec:diskmodel}. \sect{cryst_layer} defines the region for crystallization in the disk (red dashed region in \fg{schem_fig}), and \sect{dustlayer} shows the heights that dust can vertically reach (yellow line for small dust and green line for large dust in \fg{schem_fig}). Vertical mixing and radial drift are compared in \sect{mixingprocesses} to define the reservoir of crystalline dust (color-shaded regions in \fg{schem_fig}). \Sects{disklab}{MCMaxspectrum} describe modeling radial transport of dust in the midplane, and the calculated mid-infrared spectra, respectively. The outline of this modeling strategy is shown in \fg{schem_flow}.

\begin{figure}
    \centering
    \includegraphics[width=\linewidth]{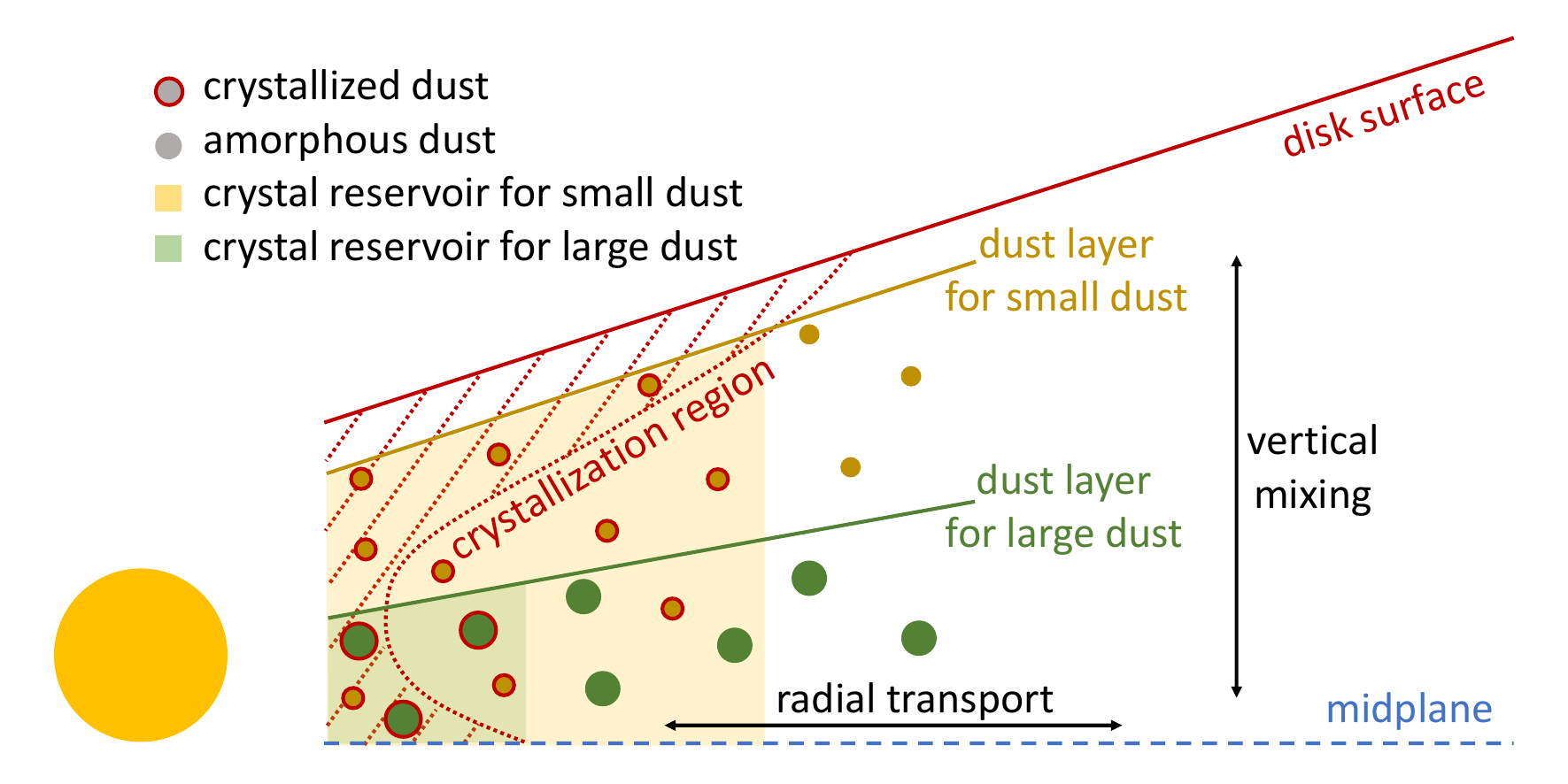}
    \caption{Sketch of modeling crystal reservoirs of different sizes of dust with crystallization and dust layers. Defining the crystallization region (red dashed region) is discussed in \sect{cryst_layer}. Because smaller dust stirs up to the upper crystallization region, the dust has a larger crystal reservoir (yellow region). The crystal reservoir for larger dust, which does not reach the upper layers, is defined by the crystallization region in the midplane (green region).}
    \label{fig:schem_fig}
\end{figure}
\begin{figure}
    \centering
    \includegraphics[width=0.7\linewidth]{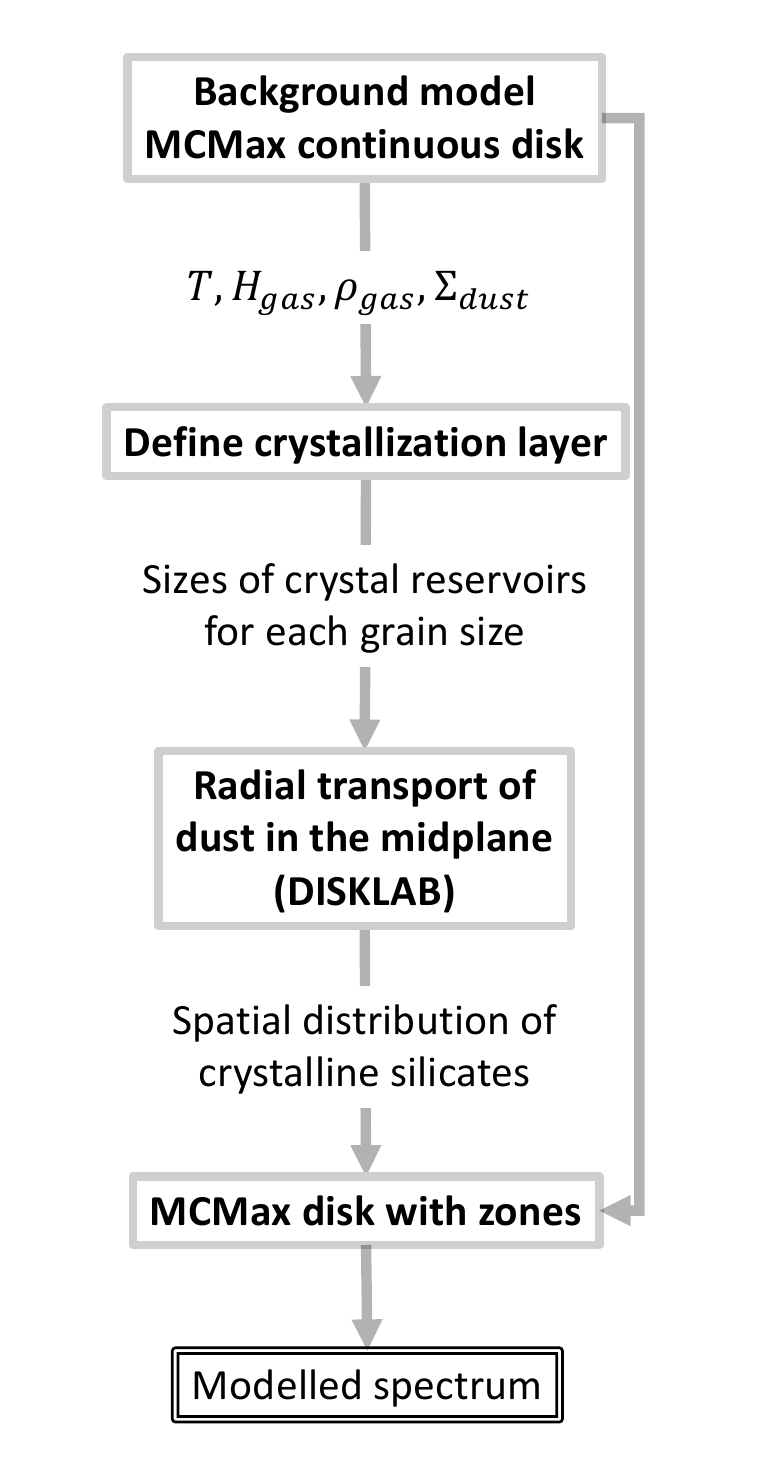}
    \caption{Schematic flow of the modeling strategy. }
    \label{fig:schem_flow}
\end{figure}

\subsection{Background disk model}
\label{sec:diskmodel}
\paragraph{}
We modeled a T-Tauri protoplanetary disk that has a power-law surface density with an exponential cutoff at $R_{\rm c} = 200$ au as
\begin{equation}
    \Sigma = \Sigma_{\rm c} \left(\dfrac{r}{R_{\rm c}}\right)^{-1} {\rm exp}\left(-\dfrac{r}{R_{\rm c}}\right),
    \label{eq:sigma1}
\end{equation}
where $\Sigma_{\rm c}$ is the characteristic surface density at 200 au with the total disk mass of 0.039 M$_{\odot}$ \citep{Klarmann_etal2018}. The scaleheight of the disk is parameterized \citep{Woitke_etal2016} with 
\begin{equation}
    H_{\rm gas} = H_{\rm c} \left( \dfrac{r}{R_{\rm c}}\right)^p,
\end{equation}
where $H_{\rm c}$ is the characteristic scaleheight of the disk model at $R_{\rm c}=200$ au as 14.14 au and $p$ is the flaring powerlaw index of the scaleheight as 1.15. 

We use MCMax, a Monte Carlo radiative transfer code \citep{Min_etal2009}, to calculate the density and temperature profiles within this T Tauri disk, as a background disk model. The properties of \text{the central} star, disk, and dust are listed in Table \ref{table:properties}. Our model is chosen to be highly turbulent with $\alpha = 10^{-2}$ to clearly see the dynamical effects on the dust distribution. The size distribution of dust follows a power-law of $a_{\rm pow}$ from the minimum size $a_{\rm min}$ to the maximum size $a_{\rm max}$ with ten size bins. The dust has a hollow volume fraction from 0 to $f_{\rm max}$ for the Distribution of Hollow Spheres \citep{Min_etal2005}. The dust in the model is composed of 93\% amorphous Mg$_2$SiO$_4$ and 7 \% amorphous carbon. We included amorphous carbon to provide opacity at optical and near-IR wavelengths, which the Mg-rich silicates hardly provide, but did not include crystalline silicates because the opacity of silicates in the UV and optical does not depend strongly on lattice structure.

\begin{table}[]
\caption{Properties of the disk model} 
\label{table:properties} 
\centering     
\begin{tabular}{cc|cc|cc}
\hline\hline
\multicolumn{2}{c|}{central star} & \multicolumn{2}{c|}{disk} & \multicolumn{2}{c}{dust} \\ \hline
$M_{\star}$ & 0.7 M$_{\odot}$ &$M_{\text{tot}}$ & 0.039 M$_{\odot}$ & $a_{\text{min}}$ & 0.05 $\mu$m \\
$L_{\star}$ & 1 L$_{\odot}$ & $\rho_{d}/\rho_{g}$ & 0.01& $a_{\text{max}}$ & 3000 $\mu$m \\
T$_{\text{eff}}$ & 4000 K & $R_{\rm in}$ & 0.07 au & $a_{\rm pow}$ & 3.5 \\
$d$ & 140 pc & $R_{\rm out}$ & 1000 au & $f_{\text{max}}$ & 0.8 \\
 &  & $\alpha$ & $10^{-2}$ & & \\
\end{tabular}
\end{table}

We used the resulting 2D temperature profile, scale height and volume density of gas, and surface density of dust from MCMax to determine a crystallization region, where conditions for crystallization by thermal annealing processes are satisfied. Note that we do not consider accretion heating in the midplane for the temperature profile even with high disk turbulence because the typical mass accretion rate of T-Tauri disks ranges from $10^{-9}-10^{-7}$ M$_{\odot}$/yr for disk turbulence of $\alpha = 10^{-2}$ and $10^{-3}$ \citep{Hartmann_etal1998}. Moreover, accretion heating increases the midplane temperature up to $500-400$ K at $0.2-1$ au \citep{Sasselov_Lecar2000,Ueda_etal2023}. If the temperature also significantly increase in the inner disk ($r < 0.2$ au) greater than 600 K by the accretion heating or low-temperature crystallization occurs, crystallinity in the inner disk would increase. Thus, our model could underestimate the crystallinity in the disk. The temperature profile and the optical depth at $10 \mu$m are shown in \fg{crystaldust_layer_tau}. Its resulting spectrum is generated by using a Monte Carlo method following \cite{Bjorkman_Wood2001} with $45^{\degree}$ inclination from the line of sight (\fg{backgroundSpec}).
\begin{figure}
    \centering
    \includegraphics[width=\linewidth]{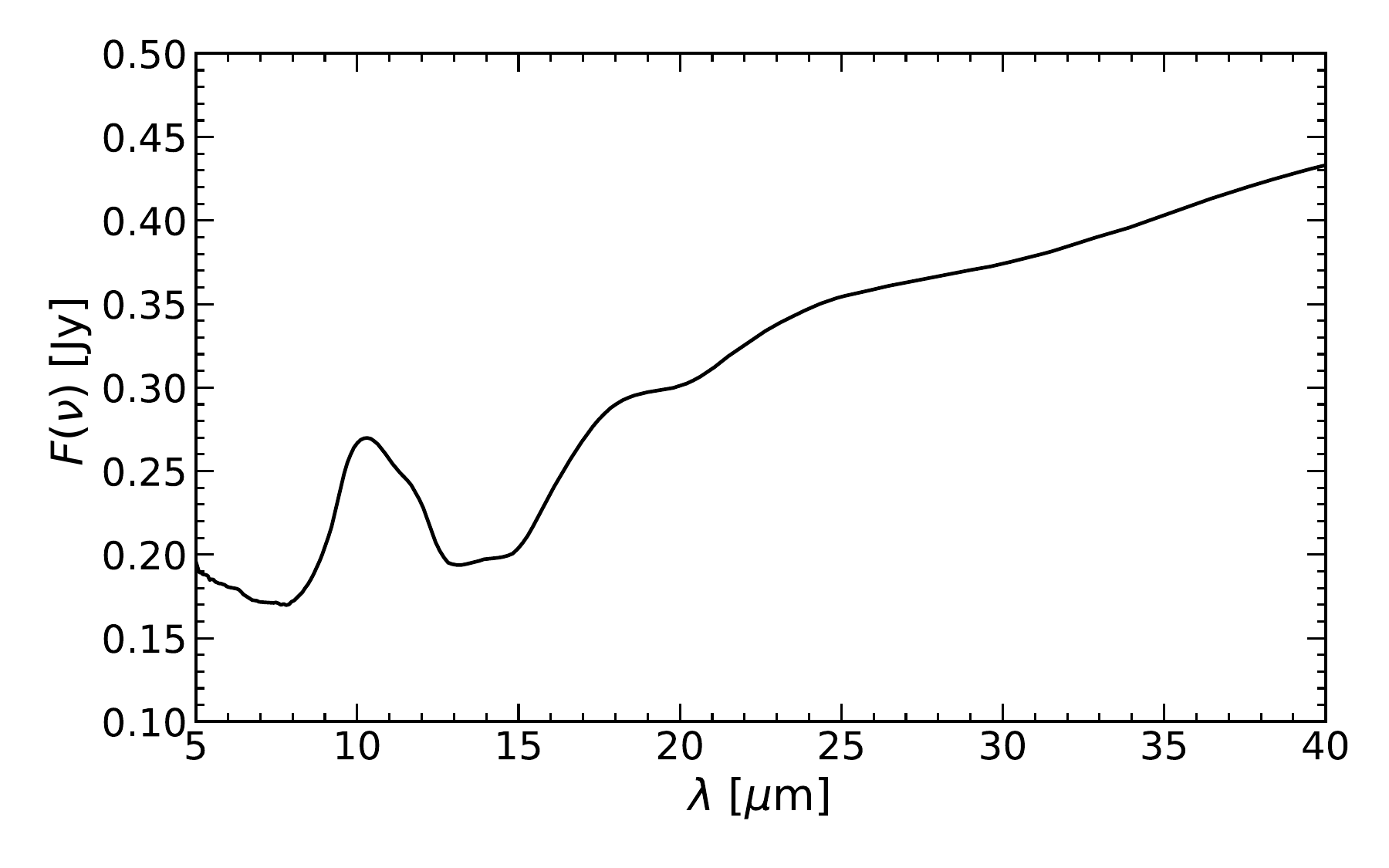}
    \caption{Modeled spectrum for the background disk model. This model only includes amorphous silicate (Mg$_2$SiO$_4$) and carbon.}
    \label{fig:backgroundSpec}
\end{figure}

\subsection{crystallization region}
\label{sec:cryst_layer}
\paragraph{}
Inner disk crystalline silicates can result from the annealing of amorphous silicates or gas phase condensation. For our study, we ignore gas phase condensation because it fully overlaps with the region where amorphous dust already has been annealed. For the crystallization, the region has to be sufficiently hot to anneal amorphous silicates on a timescale that is shorter than the timescale on which the dust is transported to a lower temperature (either vertically or radially). We name the region that satisfies this condition as the ``crystallization region.'' We compared the crystallization timescale ($\tau_{\rm cryst} $), the time required for crystallization at a given temperature, and the residence timescale ($\tau_{\rm res}$), the time that dust remains at the local height or above in the disk. Note that the crystallization timescale is grain size independent. Moreover, we do not consider partial crystallization of the dust due to the exponential dependence on temperature and micron-sized grains.

The crystallization timescale can be calculated as
\begin{equation}
    \tau_{\rm cryst } = \nu^{-1} \text{exp}(-E_{\rm a}/kT),
    \label{eq:t_crystal}
\end{equation}
where $\nu$ is the frequency of vibration of an atom or ion, and $E_{\rm a}$ is the activation energy. In this paper, we used $\nu = 2\times 10^{13}$ s$^{-1}$ as the mean vibrational frequency of the magnesium-silicate lattice \citep{Lenzuni_etal1995}. \cite{Fabian_etal2000} experimentally studied the crystallization time with given temperatures and found activation energies of their silicate samples. They suggested activation energy for forsterite (crystalline Mg$_2$SiO$_4$) of $E_{\rm a}/k = 39100 \pm 400$ K. We chose forsterite to trace the crystalline silicate in this paper because it has clearly distinguishable features at mid-infrared wavelengths. Using the temperature profile of the background disk model from MCMax, we calculated $\tau_{\rm cryst }$ and compared it to the residence timescale ($\tau_{\rm res}$). If the dust stays at its local temperature longer than $\tau_{\rm cryst}$, the dust gets crystallized.

\cite{Klarmann_etal2018} express the residence timescale as 
\begin{equation}
    \tau_{\rm res} = \left(\dfrac{H}{z_1}\right)^2 \dfrac{1}{\Omega_{\rm K} \alpha} = 1.76 {\rm yr} \left(\dfrac{z_1}{3H}\right)^{-2}\left(\dfrac{\alpha}{10^{-2}}\right)^{-1}\left(\dfrac{r}{1 {\rm au}}\right)^{3/2}\left(\dfrac{M_*}{1M_{\odot}}\right),
    \label{eq:t_res}
\end{equation}
where $H$ is the gas scale height, and $z_1$ is the lower height of the exposed layer where $\tau_r = 1$. Note that this equation is valid while the dust is still coupled to the gas. In this paper, we define $z_{1}$ differently from \cite{Klarmann_etal2018}. \Eq{t_res} calculates the residence timescale in the exposed layer. However, we set $z_1$ to be any height in the disk to calculate $\tau_{\rm res}$ at every height because we are interested in the residence timescale at its local height. If $\tau_{\rm cryst} < \tau_{\rm res}$, this region is the crystallization region and is plotted with black dots on top of our modeled temperature profile in \fg{crystaldust_layer_tau}. In the midplane, $z_{1} = 0$, so $\tau_{\rm res}$ approaches infinity. However, $\tau_{\rm cryst }$ also increases exponentially with decreasing the temperature. At 500 K, the crystallisation timescale exceeds the life time of protoplanetary disks by many orders of magnitude, inhibiting crystallization in the midplane, where the temperature is lower than 500 K. Thus, we do not consider the comparison between $\tau_{\rm cryst }$ and $\tau_{\rm res }$ in the midplane.

\begin{figure}
    \centering
    \includegraphics[width=\linewidth]{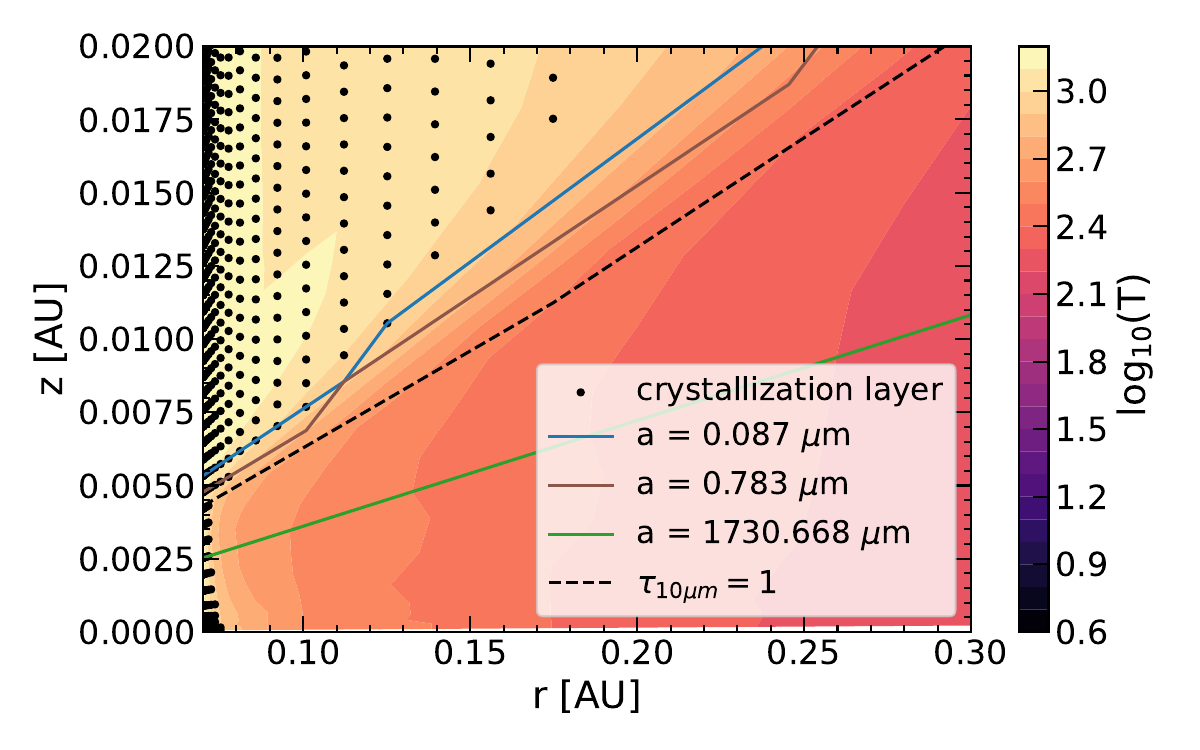}
    \caption{crystallization region and dust layers of different grain sizes with vertical optical depth $\tau_{\rm v} = 1$ at 10 $\mu$m. The temperature profile in $r$ and $z$ from MCMax is shown in the colormap, and the black dotted region is the crystallization region. Solid lines are the maximum heights of the dust layers, where the local St = $10^{-2}$ for 0.087 $\mu$m, 0.783 $\mu$m, and 1730 $\mu$m dust. The dashed line represents $\tau_{\rm v} = 1$ at 10 $\mu$m.}
    \label{fig:crystaldust_layer_tau}
\end{figure}

In the innermost disk, the midplane temperature is high enough for crystallization due to the direct radiation from the central star. Thus, for our choice of model parameters of star and disk, dust in the midplane within $\sim 0.075$ au gets fully crystallized. Beyond $\sim 0.075$ au, the crystallization region only exists in the disk surface. Dust grains must reach this upper crystallization region by vertical mixing processes in order to anneal in the outer disk. In the next sub-section, we discuss which grains reach the upper disk.

\subsection{Dust layer}
\label{sec:dustlayer}
\paragraph{}
Not all dust grains reach the upper crystallization region because of dust settling. Different sizes of dust have different Stokes number (St), which represent how the dust couples to the flow of gas as a dimensionless stopping timescale. This stopping timescale has two regimes, depending on grain size.
\begin{equation}
    t_{\rm stop} = 
    \begin{cases}
      \dfrac{\rho_{\bullet}a}{\rho_{\rm g}c_{\rm s}} & \text{for } a\leq \dfrac{9}{4}\lambda_{\rm mfp} \text{ (Epstein regime)}\\
      \dfrac{4\rho_{\bullet}a^2}{9\rho_{\rm g}c_{\rm s}\lambda_{\rm mfp}} & \text{for } a>\dfrac{9}{4}\lambda_{\rm mfp} \text{ (Stokes regime)},
    \end{cases}    
\end{equation}
where $\lambda_{\rm mfp}$ is the gas mean free path length, $\rho_{\bullet}$ is the density of the grain, $c_{\rm s}$ is the sound speed. For the $\mu$m-sized dust grains, the stopping timescale is in Epstein regime. For a given gas pressure and internal density of dust, larger grains have a greater St, making it easier to decouple from the gas at lower gas pressures. Thus, different grain sizes get decoupled from the gas at different heights. We used St to determine the height where the dust starts to settle to the midplane. For a disk with $\alpha = 10^{-2}$, dust with St $>10^{-2}$ starts to decouple from the gas in the midplane based on normalized scale height from \cite{Dubrulle_etal1995,Youdin_Lithwick2007,Birnstiel_etal2010} as $H_{\text{dust}}/H = \sqrt{\alpha / (\alpha+\text{St})}$. We calculated the height where the local St starts to exceed $10^{-2}$ for different grain sizes. In this paper, we refer to the regions from the midplane to these heights as dust layers.

In \fg{crystaldust_layer_tau}, the heights of dust layers for dust grains with size 0.087 $\mu$m, 0.783 $\mu$m, and 1730 $\mu$m are plotted with colored solid lines. As the grain size decreases, the height of the dust layer increases because smaller dust is well-coupled to the gas, while larger dust grains decouple. The dust layers overlap with the crystallization region up to 0.14 au and 0.08 au for dust grains with size 0.087 $\mu$m and 0.783 $\mu$m, respectively. 1730 $\mu$m-sized dust grain does not reach the upper crystallization region, so its crystallization is determined by the midplane temperature. The maximum distance for the crystallization in the midplane is 0.076 au, so 1730 $\mu$m-sized dust grain gets crystallized only up to 0.076 au. Beyond 0.076 au, the midplane can have an extended region for fully crystallized silicates as far as the disk surface if the vertical mixing is faster than radial drift. In that case, dust in the midplane is well mixed with the disk surface before amorphous silicates from the outer disk replenish the midplane.

\subsection{Vertical mixing and radial drift timescales}
\label{sec:mixingprocesses}
\paragraph{}
Dust in the midplane and the disk surface are mixed by disk turbulence, parameterized by $\alpha$. Amorphous silicates in the midplane reach the upper crystallization region, get crystallized, and settle back into the midplane as vertical mixing proceeds. Another dynamical process of dust is radial drift toward the central star in the midplane. If the dust gets vertically well mixed before radial drift replenishes the inner disk with amorphous silicates from the outer disk, this region is fully crystallized from the midplane to the disk surface. Thus, we compare the vertical mixing timescale  ($\tau_{\rm ver}$) and radial drift timescale ($\tau_{\rm drift}$) to find the dominant process in the inner disk. 

The vertical mixing timescale can be obtained from the turbulent diffusivity and the gas scale height \citep{Klarmann_etal2018} as
\begin{equation}
    \tau_{\rm ver} = \dfrac{H^2}{D_g} = \dfrac{1}{\Omega_{\rm K} \alpha} = 15.93 {\rm yr} \left(\dfrac{\alpha}{10^{-2}}\right)^{-1} \left(\dfrac{r}{1 \rm au}\right)^{3/2}\left(\dfrac{M_*}{1M_{\odot}}\right),
    \label{eq:t_ver}
\end{equation}
where the gas turbulent diffusivity $D_{\rm g}$ is assumed to be equal to the gas viscosity $\nu = \alpha H^2 \Omega_{\rm K}$. Note that vertical mixing can occur while the dust is coupled to the gas. The radial drift timescale is obtained from the radial dust velocity \citep{Birnstiel_etal2010}, 
\begin{equation}
    v_{\rm d} = \dfrac{1}{1+\text{St}^2}\left(v_r +\text{St}\dfrac{c^2_{\rm s}}{\Omega_{\rm K} r}\dfrac{d\ln p}{d \ln r}\right),
    \label{eq:DustRadialVel}
\end{equation}
where $p$ is the gas pressure in the midplane. $v_{\rm r}$ is the radial velocity of the gas due to the viscosity $\nu$
\begin{equation}
    v_{\rm r} = -\dfrac{3}{\sqrt r \Sigma}\dfrac{\partial(\sqrt r \Sigma \nu)}{\partial r},
    \label{eq:GasRadialVel}
\end{equation}
where $\Sigma$ is the surface density in \eq{sigma1}. Thus, the radial drift timescale is
\begin{equation}
    \tau_{\rm drift} = \dfrac{r}{v_{\rm d}}.
\end{equation}

In \fg{t_ver_rad}, $\tau_{\rm ver}$ and $\tau_{\rm drift}$ are compared over the radial distance for 0.087 $\mu$m and 1730 $\mu$m dust in disks with $\alpha = 10^{-2}, 10^{-3},$ and $10^{-4}$. $\tau_{\rm ver}$ is always shorter than $\tau_{\rm drift}$ inside 7 au, where the dust layers overlap with the upper crystallization region. Thus, this region is always fully crystallized even in the midplane, and we call this region the ``crystal reservoir.''
\begin{figure}[]
    \centering
    \includegraphics[width=\linewidth]{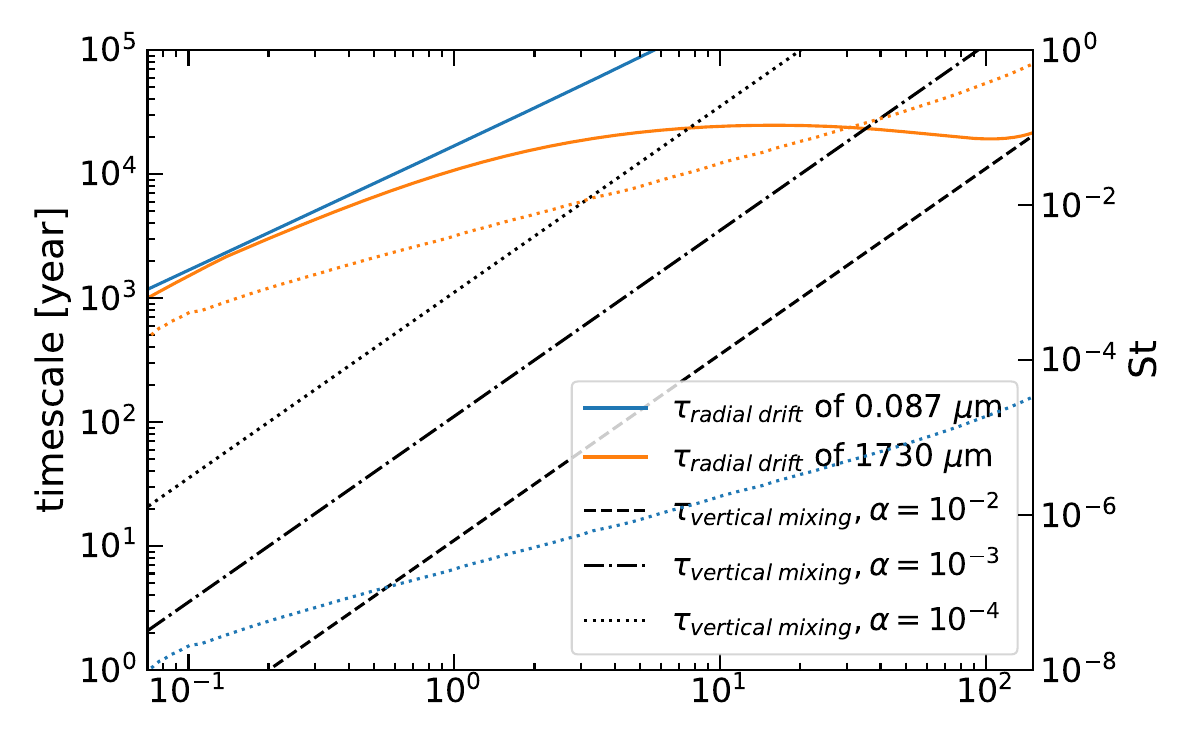}
    \caption{Radial drift and vertical mixing timescales in a disk with different disk turbulence. Blue and orange solid lines are radial drift timescales of 0.087 $\mu$m and 1730 $\mu$m dust, respectively. The colored dotted lines are St of 0.087 $\mu$m and 1730 $\mu$m dust in the midplane, following the axis on the right. St of the 1730 $\mu$m dust is larger than 0.087 $\mu$m dust, so 0.087 $\mu$m dust radially drift faster than 1730 $\mu$m dust. The dashed line is the vertical mixing for a turbulent disk with $\alpha=10^{-2}$. The dash-dotted line is for $\alpha=10^{-3}$, and the dotted line is for a quiescent disk of $\alpha=10^{-4}$.}
    \label{fig:t_ver_rad}
\end{figure}

As the crystalline silicates in the disk surface affect the midplane, the midplane also affects the disk surface. The dust in the disk experiences inward and outward dynamics. The accretion onto the central star and gas drag move the dust inward while gas density gradients diffuse the dust outward. These drift and diffusion processes affect the spatial distribution of (crystalline) dust in the midplane, and the distributed dust gets vertically mixed to the disk surface. Thus, we study how the crystalline silicates from the crystal reservoirs get radially distributed in the midplane and whether the distribution represents the disk surface by using the DISKLAB code.  

\subsection{Radial transport process}
\label{sec:disklab}
\paragraph{}
The DISKLAB code\footnote{The code is not yet published, but access to the code is available by request to Cornelis Dullemond (dullemond@uni-heidelberg.de).} solves the time-dependent evolution of a viscous disk together with the radial transport of dust in the midplane. In this code, the gaseous disk viscously evolves by a single time step $\Delta t$ with an implicit integration method. Based on the computed evolution of the gaseous disk, the code calculates the dynamics of the dust, taking into account inward gas-drag drift and accretion and outward diffusion along with the gas. We used DISKLAB to study how crystalline silicates in the crystal reservoir get radially transported within the lifetime of a protoplanetary disk. 

The time evolution of the gas surface density is expressed as 
\begin{equation}
    \dfrac{\partial\Sigma}{\partial t}+\dfrac{1}{r}\dfrac{\partial}{\partial r}(r\Sigma v_{\rm r}) = 0,
    \label{eq:DiskEvol}
\end{equation}
where $v_{\rm r}$ is the radial velocity due to the viscosity in \eq{GasRadialVel}, and we neglect gas accretion from a disk envelope because the amount of possible in-falling gas is small in Class II disks. Thus, \eq{DiskEvol} becomes 
\begin{equation}
    \dfrac{\partial \Sigma}{\partial t} - \dfrac{3}{r} \dfrac{\partial}{\partial r} \left(\sqrt r \dfrac{\partial(\sqrt r \Sigma \nu)}{\partial r}\right) = 0.
    \label{eq:FinalDiskEvol}
\end{equation}

In addition to the evolution of the viscous gas disk, the code calculates the radial transport process of dust. The time-dependent equation of dust drift and diffusion follows \cite{Birnstiel_etal2010} as
\begin{equation}
    \dfrac{\partial \Sigma_{\rm d}}{\partial t} + \dfrac{1}{r}\dfrac{\partial (r\Sigma_{\rm d} v_{\rm d})}{\partial r} - \dfrac{1}{r}\dfrac{\partial}{\partial r}\left(rD_{\rm d}\Sigma \dfrac{\partial}{\partial r}\left(\dfrac{\Sigma_{\rm d}}{\Sigma}\right)\right) = 0,
    \label{eq:DustRadialMixing}
\end{equation}
where $v_d$ is the radial dust velocity calculted as \eq{DustRadialVel}. The diffusion coefficient $D_d$ is 
\begin{equation}
    D_{\rm d} = \dfrac{1}{\text{Sc}}\dfrac{1}{1+\text{St}^2}\nu,
\end{equation}
where Sc is the Schmidt number of gas, the ratio of gas turbulent viscosity $\nu$ over gas turbulent diffusivity $D_{\rm g}$, Sc $= \nu/D_{\rm g}$. In different literature studies, the Schmidt number varies from 0.3 to 10 in numerical simulations of turbulent protoplanetary disks \citep{Carballido_etal2005,Johansen_Klahr2005, Johansen_etal2006, Turner_etal2006}. For simplicity, we set the fiducial value for the Schmidt number as 1 in the DISKLAB code, a value generally adopted for radial transport modeling \citep{Gail2001,Wehrstedt_Gail2002,Ilgner_etal2004}.

The DISKLAB code solves \eq{DustRadialMixing} and provides the dust surface density for each grain size as a function of time independently. Thus, we obtain the abundance of crystalline silicates for each grain size as $\Sigma_{\text{cryst}, a}/\Sigma_{\text{dust},a}$, where $\Sigma_{\text{cryst}, a}$ is the surface density of crystalline dust with size $a$, and $\Sigma_{\text{dust},a}$ is the surface density of total dust with size $a$.

In the DISKLAB model, we used the same disk surface density as \eq{sigma1} with $R_{\rm c} = 70$ au instead of 200 au to avoid the pile-up effect at the closed end of the disk model at 1000 au. $\Sigma_{\rm c}$ was adjusted to make the surface density of the inner disk $< 70$ au the same as the background disk model in \sect{diskmodel}. Properties of the star, disk, and dust also follow the background disk model (Table \ref{table:properties}). 

As initial conditions, we defined the outer edges of crystal reservoirs ($r_{\rm cr}$) for each grain size. For distances smaller than $r_{\rm cr}$, the crystallinity ($\Sigma_{\rm cryst}/\Sigma_{\rm dust}$) is 100 \%. At distances larger than $r_{\rm cr}$, we allow the disk to initially have a constant crystallinity, which is chosen to be 3 \% crystallinity in the fiducial model. This outer disk initial crystallinity is introduced to take into account any early formation and distribution of crystalline silicates, such as thermal annealing in an early stage of disk formation \citep{Dullemond_etal2006}, lightning in a nebula \citep{Desch_Cuzzi2000}, and nebular shocks \citep{Harker_Desch2002}. 

The left panel of \fg{radial_dist_all} shows how crystalline silicates from the crystal reservoirs are radially distributed after 1 Myr, a typical evolution timescale for the protoplanetary disk, for 0.087 $\mu$m (solid line), 0.26$\mu$m (dashed-dotted line), 2.35 $\mu$m (dashed line) and 1730 $\mu$m (dotted line) dust. The region where the crystallinity $\Sigma_{\text{cryst}}/\Sigma_{\text{dust}} = 1$  in the innermost disk is the crystal reservoir. In \Fg{radial_dist_all}(a), 0.087 $\mu$m dust reaches 10 \% crystallinity at $\sim$ 0.8 au after 1 Myr while 1730 $\mu$m dust reaches 10 \% at 0.3 au. The grain sizes in \fg{radial_dist_all} are selected to illustrate different distributions of dust. 

\begin{figure*}
    \centering
    \includegraphics[height=0.9\textheight]{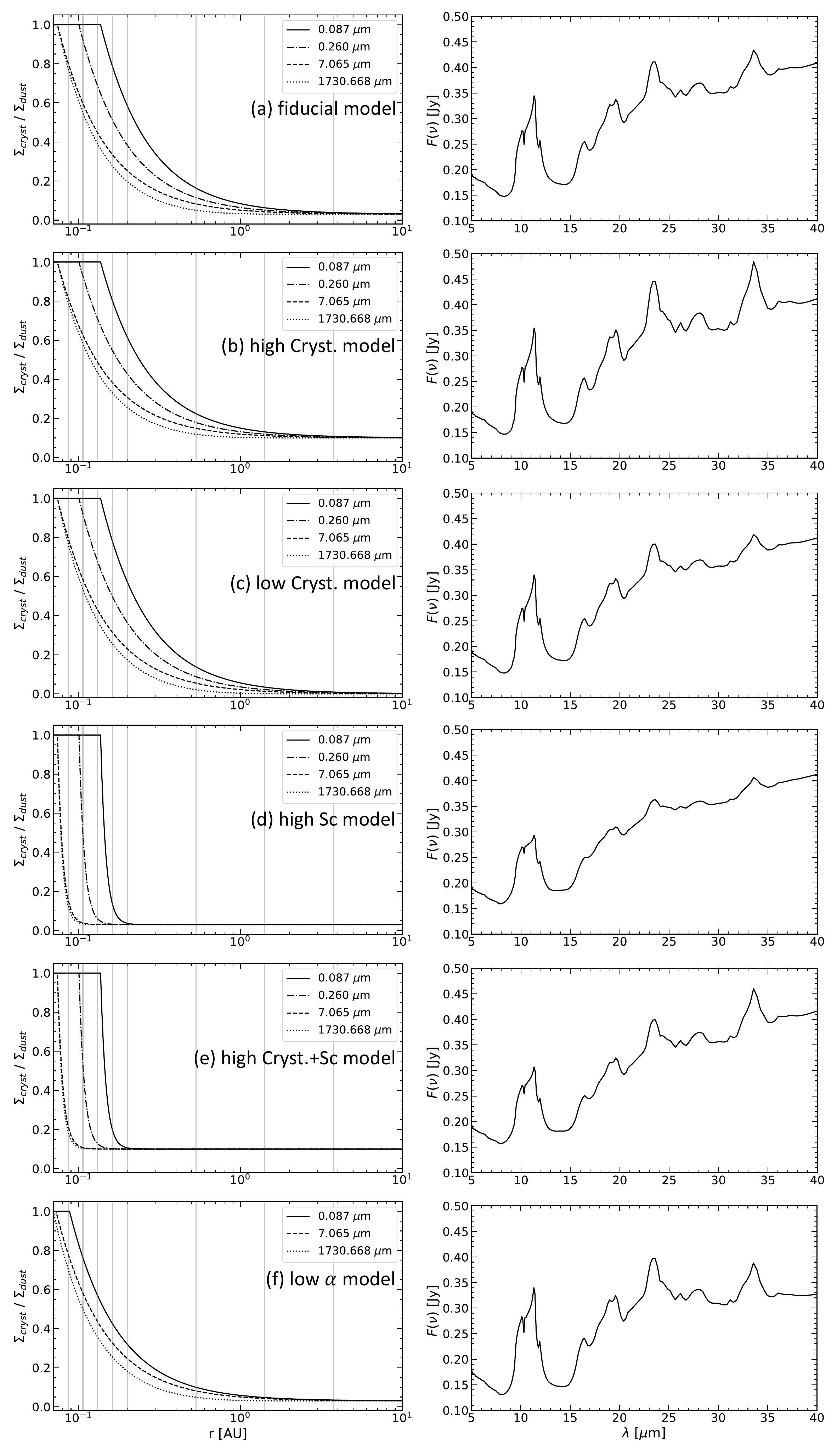}
    \caption{Left panels: Spatial distributions of crystalline silicates with different parameters as a result of radial transport processes after 1 Myr. Crystalline silicates from the crystal reservoir (horizontal region in the inner disk) diffuse and approach the initial crystallinity in the outer disk. The crystallinity was estimated from the surface density of crystalline silicates divided by the surface density of total dust ($\Sigma_{\rm cryst}/\Sigma_{\rm dust}$). Right panels: infrared spectra of the different disk models.}
    \label{fig:radial_dist_all}
\end{figure*}

The distance to which the dust can diffuse outwards is still in the region where vertical mixing is more effective than radial drift in \Fg{t_ver_rad}. For example, 0.087 $\mu$m dust takes about 1 Myr to diffuse out to 1 au, where $\tau_{\rm ver} < \tau_{\rm drift}$. Thus, the spatial distribution of the crystalline silicates in the midplane still represents the distribution in the disk surface. This distribution was applied to MCMax to model the disk's mid-infrared spectrum.

\subsection{Model infrared spectra}
\label{sec:MCMaxspectrum}
\paragraph{}
We divided the spatial distribution of crystalline silicates resulting from DISKLAB into ten radial zones that sample their abundance gradients; the zones are shown in \fg{radial_dist_all}(a) with vertical gray lines. The radial zones are applied to MCMax with the mean crystallinity of each zone, taken from DISKLAB at 1 Myr. The location of each zone is listed in Table \ref{table:zones}. In MCMax, we introduce two grain compositions, that we refer to as ``amorphous'' and ``crystalline''. The amorphous component has 93\% amorphous Mg$_2$SiO$_4$ with 7\% amorphous carbon, and the crystalline component has 93\% forsterite (crystalline Mg$_2$SiO$_4$) with 7\% amorphous carbon. All other parameters of the disk follow the background disk model (see \sect{diskmodel}). The mass abundance of crystalline and amorphous grains of size $a$ is estimated based on the mean crystallinity of each zone and the grain size distribution. The grain size distribution follows a powerlaw
\begin{equation}
    \dv{n}{a} \propto a^{-a_{\rm pow}},
\end{equation}
which results in 
\begin{equation}
    n(a) = n_0\cdot a^{-a_{\rm pow}+1},
\label{eq:sizedist}
\end{equation}
where $n_0$ is constrained by
\begin{equation}
    M_{\rm dust} = \dfrac{4}{3} \pi \rho_{\bullet} n_0 \sum_{a_i = a_{\min}}^{a_{max}} a_{i}^{-a_{\rm pow}+4},
\label{eq:const}
\end{equation}
where $\rho_{\bullet}$ is the density of the grain. Thus, the total mass of grains of size $a$ is $M_a = m_a\cdot n(a)$, where $m_a$ is the mass of a single dust grain. The constant in \eq{sizedist} can be estimated using the total dust mass. Note that our model ignores the evolution of the total dust mass; this is reasonable because of the low mass accretion rates we assume and the subtle effect of dust evolution on modeled spectra in the mid-infrared wavelength \citep{Dullemond_Dominik2004a,Greenwood_etal2019}. Although the dust surface density evolves over time due to the short radial drift timescale of large dust grains, the evolution of the smallest grains is not significant, and these smallest grains dominate the mid-infrared spectrum. Therefore, the mass abundances of crystalline dust of grain size $a$ in a zone is
\begin{equation}
    \left(\dfrac{M_{a,{\rm cryst}}}{M_{\rm dust}}\right)_{\rm zone} =  \dfrac{M_a}{M_{\rm dust}} \cdot \left(\dfrac{\Sigma_{\rm cryst}}{\Sigma_{\rm dust}}\right)_{\rm zone},
\end{equation}
where the second term on the right-hand side is the mean crystallinity of each zone from DISKLAB. The mass abundance of amorphous silicate is 
\begin{equation}
    \left(\dfrac{M_{a,{\rm amorp}}}{M_{\rm dust}}\right)_{\rm zone} =  1 - \left(\dfrac{M_{a,{\rm cryst}}}{M_{\rm dust}}\right)_{\rm zone}.
\end{equation}
As a result, the modeled spectrum is shown on the right panel of \fg{radial_dist_all}. Crystalline silicate features are present at wavelengths around 11, 16, 19, 23, 28, and 33 $\mu$m.
\begin{table}[]
\caption{Radial ranges of zones in a disk} 
\label{table:zones} 
\centering     
\begin{tabular}{ccc}
\hline\hline
zone \# & $r_{\rm in}$ [au] & $r_{\rm out}$ [au] \\ \hline
zone 1 & 0.07 & 0.086 \\
zone 2 & 0.086 & 0.107 \\
zone 3 & 0.107 & 0.131 \\
zone 4 & 0.131 & 0.162 \\
zone 5 & 0.162 & 0.2 \\
zone 6 & 0.2 & 0.532 \\
zone 7 & 0.532 & 1.414 \\
zone 8 & 1.414 & 3.761 \\
zone 9 & 3.761 & 10 \\
zone 10 & 10 & 1000 \\ \hline
\end{tabular}
\end{table}

\section{Effect of the spatial distribution on mid-infrared spectrum}
\label{sec:dist2spec}
\paragraph{}
Strong features of forsterite appear at various mid-infrared wavelengths. The features qualitatively agree with observed Spitzer spectra in e.g. \cite{ KesslerSilacci_etal2006}, \cite{Furlan_etal2006}, and \cite{Watson_etal2009} except those at $\sim 10 \mu$m (discussed in \sect{qual_comparison}). To investigate how the spatial distribution of forsterite dust affects these features, we first investigated how different zones in the disk contribute to the emission features. 

\subsection{The contributions of disk zones to dust features}
We have calculated the fractional contribution to the crystalline silicate features as a function of disk zone, using the cumulative spectrum as a function of a number of zones. An example of such cumulative spectra is given in the upper panel of \fg{contribution_fidu}. We modeled ten spectra with MCMax by stacking the zones one by one. For example, the first model has the first innermost zone (0.07 - 0.09 au) of the fiducial model, and the second model has the first two inner zones (0.07 - 0.11 au). The continuum was defined with a linear fit between the start of a feature($\lambda_1$) and the end of the feature($\lambda_2$), and $\lambda_1$ and $\lambda_2$ are selected based on Table \ref{table:wave_totflux}. We measured the ratios of total flux for each model, from zone 1 to zone $i$, and the fiducial model, from zone 1 to zone 10, at each feature, $F_{\text{feature}, i}/F_{\text{feature}, 10}$. The contribution of the second inner zone (0.09 - 0.11 au) was estimated by subtracting the ratios of the first model, $F_{\text{feature}, 2}/F_{\rm feature,10}$ - $F_{\text{feature}, 1}/F_{\rm feature,10}$. Note that the $10 \mu$m feature covers both amorphous silicates at 9.8 $\mu$m and the forsterite feature at 11.3 $\mu$m because we lumped the feature by integrating from 9 to 13 $\mu$m. Thus, the calculation for the contribution of disk zones for $10 \mu$m feature is for all silicates in the disk while other features are dominated by forsterite in the disk.
\begin{figure}
    \centering
    \includegraphics[width=\linewidth]{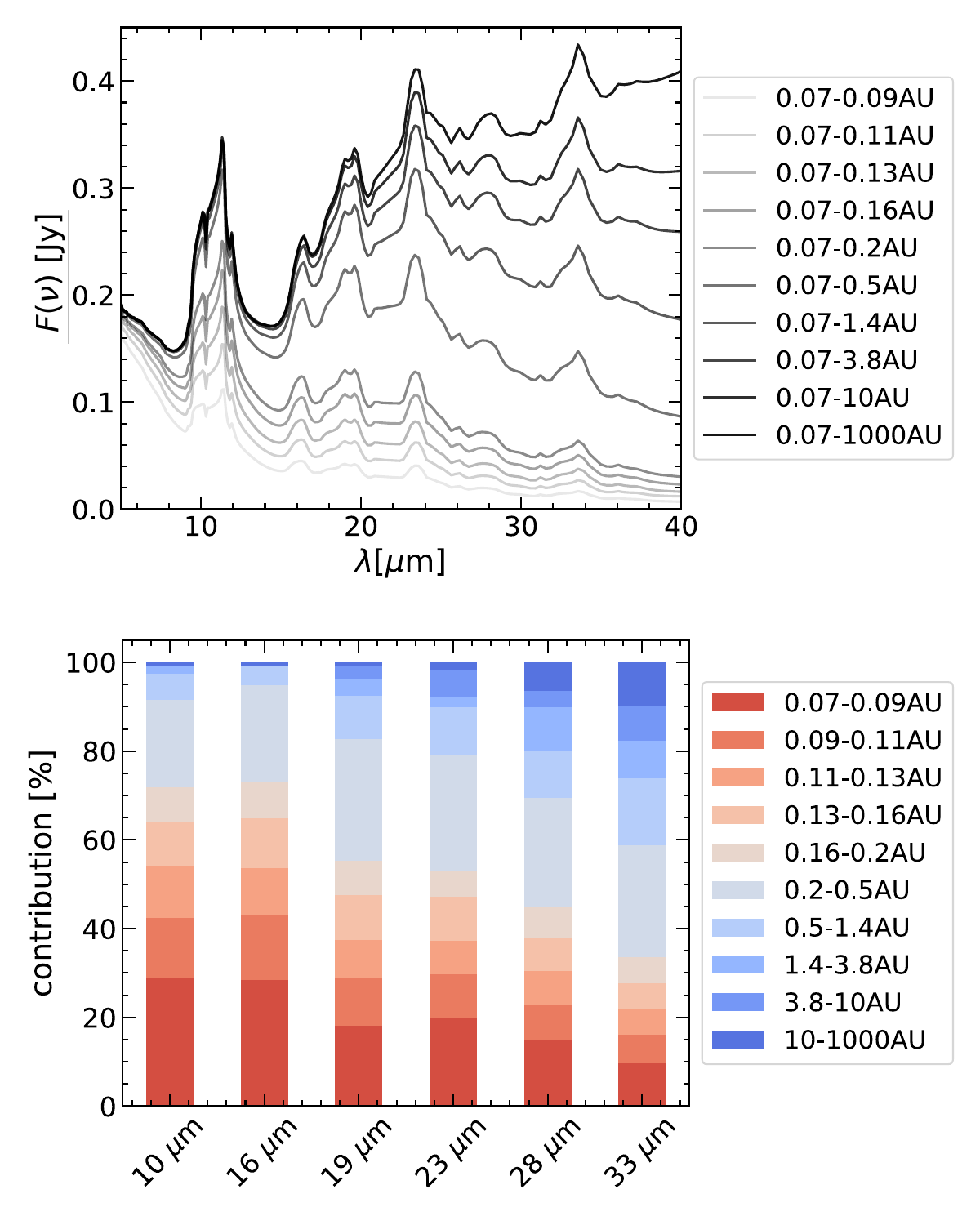}
    \caption{\textit{Top}: Cumulative spectra of the fiducial model. \textit{Bottom}: Contributions of disk zones of the fiducial model to forsterite features at 10 $\mu$m, 16 $\mu$m, 19 $\mu$m, 23 $\mu$m, 28$\mu$m, and 33 $\mu$m.  }
    \label{fig:contribution_fidu}
\end{figure}

\begin{table}[]
\caption{Wavelength ranges used to measure the total flux for each forsterite feature.} 
\label{table:wave_totflux} 
\centering  
\begin{tabular}{ccc}
\hline\hline
feature & $\lambda_1$[$\mu$m]  & $\lambda_2$ [$\mu$m] \\ \hline
10 $\mu$m & 9 & 13 \\
16 $\mu$m & 15.3 & 17.2 \\
19 $\mu$m & 17.2 & 20.5 \\
23 $\mu$m & 22.5 & 24 \\
28 $\mu$m & 26.7 & 29 \\
33 $\mu$m & 31.7 & 35 \\ \hline
\end{tabular}
\end{table}

The lower panel of \fg{contribution_fidu} shows how much each zone contributes to generating the features around 10 $\mu$m, 16 $\mu$m, 19 $\mu$m, 23 $\mu$m, 28 $\mu$m, and 33 $\mu$m. As expected, the inner disk contributes more to shorter wavelength features, and the outer disk more to longer wavelength features. For example, the innermost disk (0.07-0.09 au) contributes $\sim$ 30\% to the 10 $\mu$m feature and $\sim$ 10\% to the 33 $\mu$m feature. The outermost disk (10-1000 au) contributes less than 1 \% to the 10 $\mu$m feature and $\sim$ 10 \% to the 33 $\mu$m feature. Half light radii for each feature are $0.13$ au for the 10 $\mu$m and 16 $\mu$m features, 0.16-0.2 au for the 19 $\mu$m and 23 $\mu$m features, and 0.2-0.5 au for the 28 $\mu$m and 33 $\mu$m features. 

This result does not change if the crystallinity is the same across the entire disk. \Fg{contri_10Cryst} shows the contributions of each disk zone similar to \fg{contribution_fidu}, but for a disk with 10 \% constant crystallinity across the entire disk. Thus, the large contribution of the inner disk to the shorter wavelength is not just due to the high crystallinity of the inner disk in the fiducial model. In this model, the contribution of the inner disk drops more quickly at longer wavelength features, and the outer disk contributes more compared to the fiducial model. This is because the inner and outer disks have the same abundance of crystalline silicates while the fiducial model has a much higher abundance of crystalline silicates in the inner disk. Overall, the trend of the contribution of zones to each feature remains the same as in the fiducial model. In both cases, the contribution of the outer disk to the 10 $\mu$m and 16 $\mu$m features are very small, so the strength of the features becomes sensitive to the details of the continuum subtraction. Thus, we combined the contributions of outer disks $< 1 \%$ for the 10 $\mu$m and 16 $\mu$m features.

\begin{figure}[h]
    \centering
    \includegraphics[width=\linewidth]{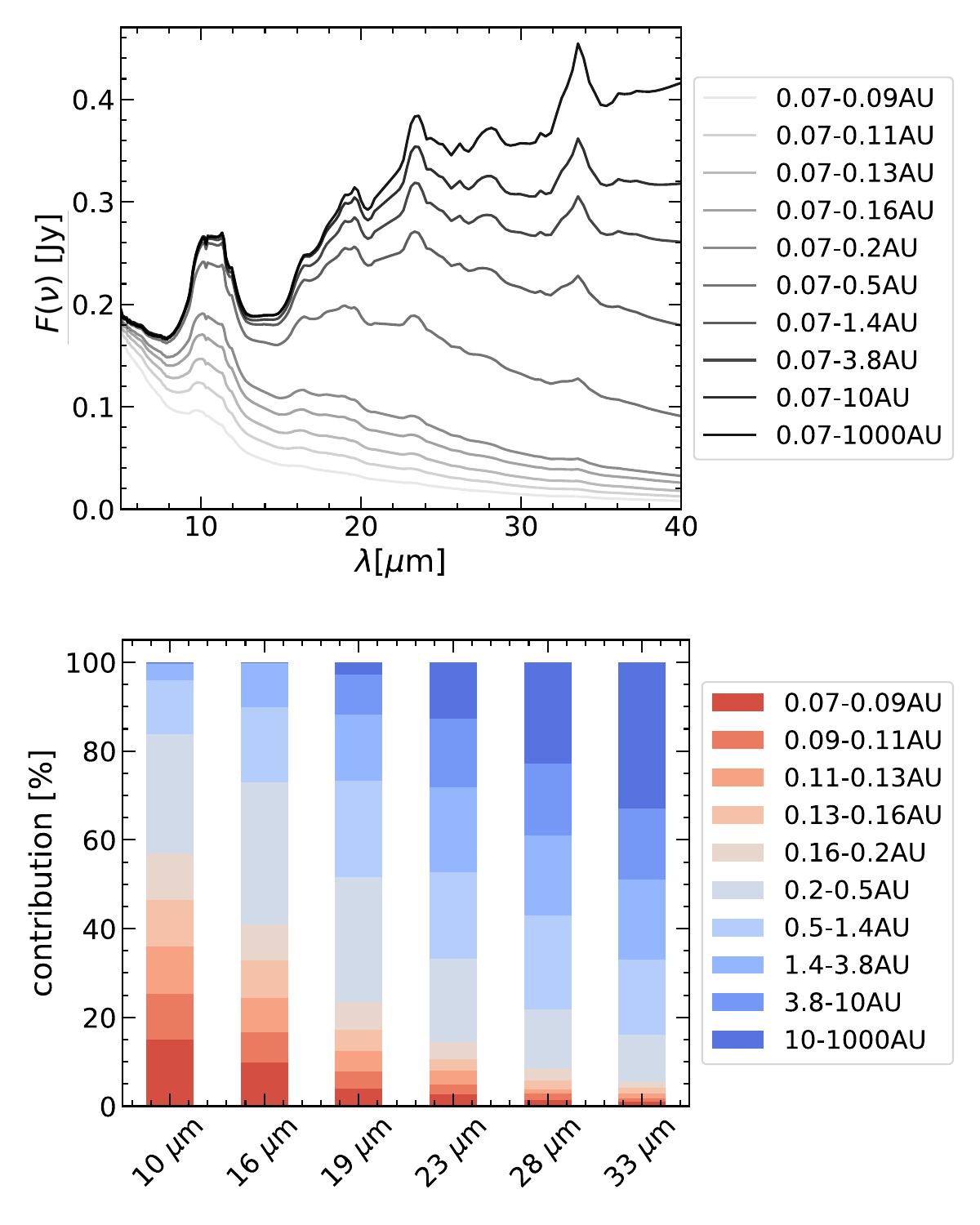}
    \caption{\textit{Top}: Cumulative spectra of the 10 \% crystalline disk model. \textit{Bottom}: Contribution of disk zones of the 10 \% crystalline disk model to forsterite features. }
    \label{fig:contri_10Cryst}
\end{figure}

Because the relative contributions of different parts of the disk are different for each feature, the spatial distribution of forsterite affects the strengths of features. In the next section, we will study how the parameters change the distribution of forsterite and how the different distributions affect the modeled spectrum. 

\subsection{Variation of the spatial distribution}
\label{sec:parameter}
\paragraph{}
The main model parameters that affect the spatial distribution of crystalline silicates are the turbulence parameter $\alpha$, the Schmidt number Sc, and the initial crystallinity. $\alpha$ changes the disk properties, and the different disk properties define new $r_{\rm cr}$. Thus, we generated a new background disk model with MCMax as \sect{diskmodel}. Sc changes how much the dust gets diffused to the outer disk along with the gas, so it determines how far the crystalline silicates can diffuse to the outer disk \citep{Clarke_Pringle1988, Pavlyuchenkov_Dullemond2007}. The initial crystallinity changes the crystallinity beyond the crystal reservoir, so the radially transported crystalline silicates approach different outer disk crystallinity. The parameter space is summarized in Table \ref{table:params}.

\begin{table}[]
\caption{Model setups for parameters to vary the spatial distribution.} 
\label{table:params} 
\centering     
\begin{tabular}{lccc}
\hline\hline   
\multicolumn{1}{c}{Description} & \begin{tabular}[c]{@{}c@{}}initial \\ crystallinity\\ {[}\%{]}\end{tabular} & \begin{tabular}[c]{@{}c@{}}Schmidt\\ number\\ (Sc)\end{tabular} & \begin{tabular}[c]{@{}c@{}}disk\\ turbulence\\ ($\alpha$)\end{tabular} \\ \hline
1. fiducial & 3 & 1 & $10^{-2}$  \\
2. high Cryst. & 10 & 1 & $10^{-2}$  \\
3. low Cryst. & 0 & 1 & $10^{-2}$  \\
4. high Sc & 3 & 10 & $10^{-2}$  \\
5. high Cryst.+Sc & 10 & 10 & $10^{-2}$  \\
6. low $\alpha$ & 3 & 1 & $10^{-4}$  \\ \hline
\end{tabular}
\end{table}

\begin{figure}
    \centering
    \includegraphics[width=\linewidth]{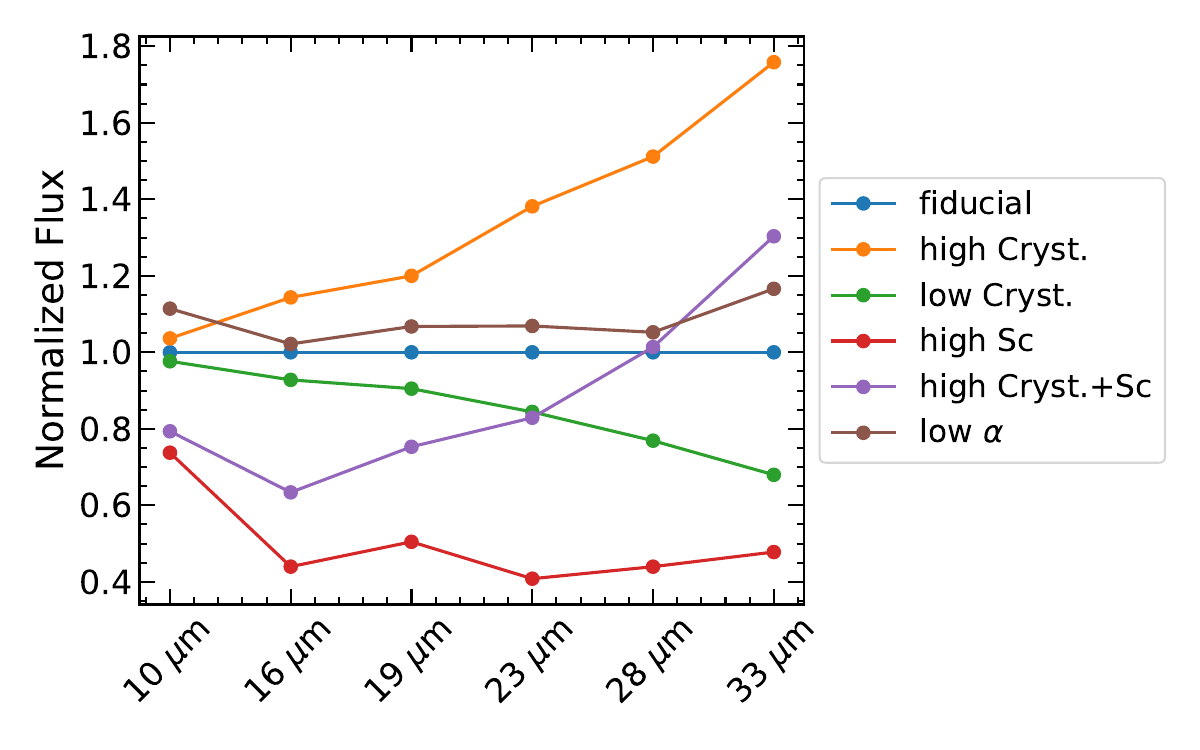}
    \caption{Normalized dust feature strength in models with different parameters as indicated in legend.} 
    \label{fig:parameter}
\end{figure}

First, we increased the initial crystallinity from the fiducial value of 3 \% to 10 \% (high Cryst.). In \fg{radial_dist_all}b, the crystallinity in the crystal reservoir is still 100 \%, and the crystallinity beyond $r_{\rm cr}$ is enhanced. Because the outer disk contributes more to longer wavelengths, the forsterite features get stronger at longer wavelengths (orange line in \fg{parameter}). On the other hand, a lower initial crystallinity of  0 \% (low Cryst.) behaves opposite to the high Cryst. model. The features get weaker at longer wavelengths (green line in \fg{parameter}). 

We also changed the Sc value to 10 (high Sc). This means that dust diffusivity is 10 times lower than gas diffusivity, so the dust has difficulty diffusing to the outer disk. As a result, crystalline silicates do not diffuse outward as much as in the fiducial model. Thus, the disk beyond $r_{\rm cr}$ has a lower crystallinity than the fiducial model and quickly converges to the initial crystallinity of the outer disk, as shown in \fg{radial_dist_all}d. In general, the features are weaker than in the fiducial model. The less-crystalline inner disk beyond $r_{\rm cr}$ still affects longer wavelengths, so the 33 $\mu$m feature also weakens, even with the same crystallinity of the outer disk as in the fiducial model (red line in \fg{parameter}). Moreover, the $ 10 \mu$m feature weakens less than other features because it is mostly affected by the innermost disk, where the crystal reservoirs are. We do not consider Sc $<$ 1 because this means the dust diffuses outward faster than the gas, and our size distribution of dust cannot diffuse faster than the gas.

Moreover, we applied high initial crystallinity together with high Sc (high Cryst.+Sc). The overall crystallinity right after $r_{\rm cr}$ is less than the fiducial model, but the outer disk has a higher initial crystallinity as shown in \fg{radial_dist_all}e. Thus, the features at shorter wavelengths are weaker than the fiducial model, following the high Sc model. On the other hand, 28 $\mu$m and 33 $\mu$m features are stronger, following the high crystallinity model (purple line in \fg{parameter}).

\begin{figure}
    \centering
    \includegraphics[width=\linewidth]{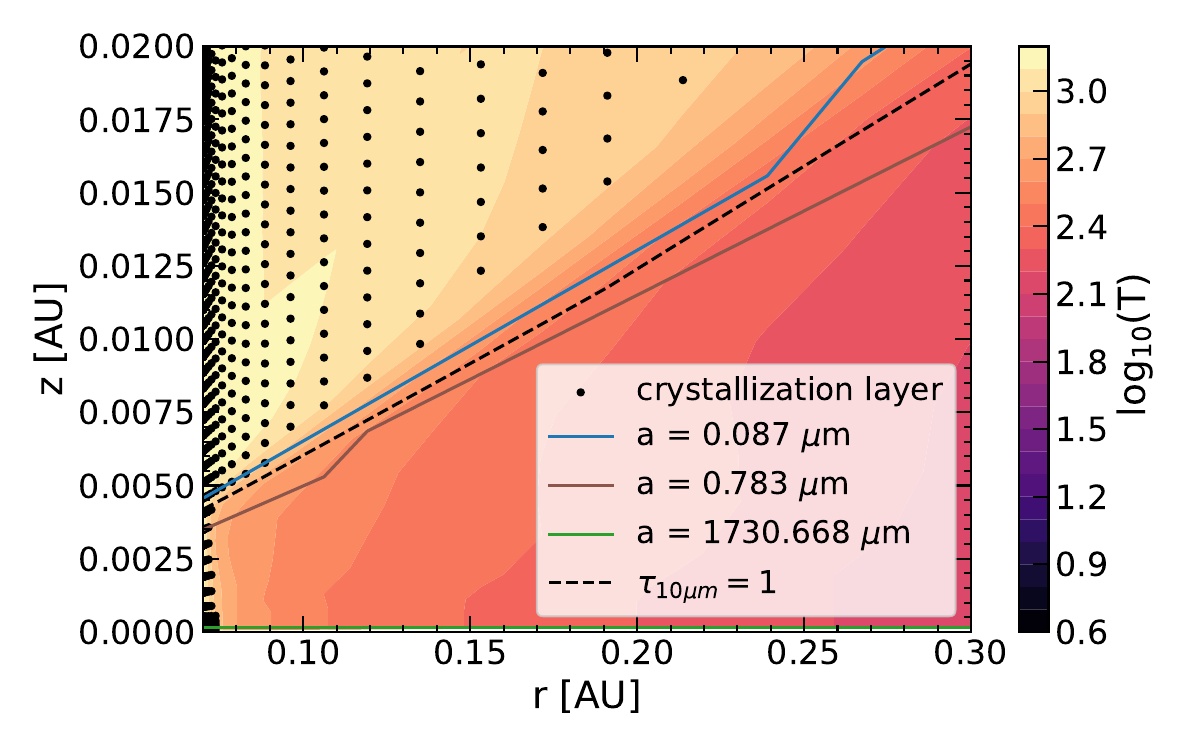}
    \caption{Crystallization region, the dust layers of different grain sizes, and the vertical optical depth $\tau_{\rm v} = 1$ at $10 \mu$m in a quiescent disk, $\alpha = 10^{-4}$.} 
    \label{fig:lowturbdisk}
\end{figure}
For the low $\alpha$ model, we generated a new disk model with a disk turbulence of $\alpha=10^{-4}$. Because the disk is more quiescent, larger dust grains cannot be stirred up as much as in the fiducial model by vertical mixing. The dust opacity in the upper disk decreases, and the layer where $\tau=1$ is reached at a lower height. Thus, the disk gets warmer close to the midplane. The temperature structure of this disk is shown in \fg{lowturbdisk}. Dust does not reach as high as in the fiducial model because the dust settles to the midplane due to the effective decoupling from the gas. As a result, the dust layers that overlap with the upper crystallization region are smaller than in the fiducial model ($r_{\rm cr} \sim 0.9$ au for $0.087 \mu$m dust). Despite the smaller crystal reservoirs, the smallest dust grains ($a = 0.087 \mu$m) diffuse outward as much as in the fiducial model (\fg{radial_dist_all}f), and these smallest dust grains are more dominant in the disk surface than in the fiducial model. Therefore, the feature strengths become slightly stronger (brown line in \fg{parameter}).

In all models, the forsterite features in the 10 $\mu$m wavelength range are very sharp while the background model spectrum in \fg{backgroundSpec} is more comparable to observations. However, the background model, which only contains amorphous silicates, is still not physical because protoplanetary disks contain some fraction of crystalline silicates due to the high temperature of the inner disk. We will qualitatively compare all our models, including the background model, with observations and suggest appropriate approaches to make our models more comparable with observations in \sect{qual_comparison}.

\section{Qualitative comparison with observations}
\label{sec:qual_comparison}
\paragraph{}
In this section, we compare our model spectra with spectroscopic infrared spectra. We follow a method to calculate the strengths and shapes of silicate features introduced in \cite{vanBoekel_etal2003} and \cite{Olofsson_etal2009}. The feature strength at certain wavelength $x$ $\mu$m is measured as 
\begin{equation}
    \text{F}x = 1 + (f_{x\;\mu\rm m, cs} / <f_{\rm c}>),
\end{equation}
where $f_{x\;\mu\rm m,cs}$ is linear continuum subtracted flux at $x$ $\mu$m, and $<f_{\rm c}>$ is the mean of the linear continuum. From the ratio of amorphous feature strength and crystalline feature strength, we can also measure the shape of the silicate feature. Based on this comparison, we discuss the strong 10 $\mu$m feature in our model spectra as shown in \fg{radial_dist_all} and suggest possible scenarios based on an extended parameter study to resolve the difference between our model and observation in \sect{ext_param}.

\subsection{The 23 $\mu$m and 10 $\mu$m forsterite feature strengths.}
\label{23_10feature}
\paragraph{}
Observed infrared spectra of planet-forming disks show a large range of shapes, resulting from a combination of different disk geometries, grain sizes in the disk surface layers, and grain compositions (see references in \sect{intro}). Our study is not aimed at reproducing this spectral richness. However, our fiducial model as well as the parameter study models all show disk spectra with distinct crystalline silicate features (\fg{radial_dist_all}). Clearly, our models represent a subset of the observed disks, that show crystalline silicate features similar in strength to our model spectra.

We quantify this using the 23 $\mu$m forsterite feature, which is relatively strong and can be easily separated from the contribution of amorphous silicates. We measure the crystalline feature at 23 $\mu$m as 
$\text{F}23 = 1 + (f_{23\mu\rm m, cs} / <f_{\rm c}>)$. We also measure the ratio between 24 $\mu$m strength and 23 $\mu$m strength (F24/F23), following \cite{Olofsson_etal2009}. A linear continuum was selected from 22.5 $\mu$m to 24.2 $\mu$m. In our fiducial model, F23 = 1.15 and F24/F23 = 0.94, which is in the same range as observed in T-Tauri disks (see Figure 5 in \cite{Olofsson_etal2009}). The 23 $\mu$m feature is dominated by emission from regions beyond 0.2 au in \fg{contri_10Cryst}, or half of the emission comes from the region in \fg{contribution_fidu}. Thus, the modeled distribution of crystalline silicates on these spatial scales seems compatible with observations.

However, the 10 $\mu$m feature shape is not what is typically observed. In the fiducial model and its variations within the parameter space, it is strongly dominated by crystalline silicates, but many analyses of observed 10 $\mu$m features show that this feature is dominated by amorphous silicates, with crystallinity fractions of a few percent up to 20 percent \citep{Apai_etal2005}. \Fg{obser_10micron} shows the ratio between 9.8 $\mu$m strength and 11.3 $\mu$m strength (F9.8/F11.3) over 9.8 $\mu$m strength (F9.8) of our models and compares this to observations of a sample of T-Tauri disks and pre-transitional disks in Combined Atlas of Sources with Spitzer IRS Spectra (CASSIS) database \citep{Lebouteiller_etal2011}. We measure the strengths of the 9.8 $\mu$m and 11.3 $\mu$m features as F9.8$= 1 + (f_{9.8,\rm cs} / <f_{\rm c}>)$, following \cite{vanBoekel_etal2003} and \cite{Olofsson_etal2009}. Spitzer IRS observations show band strengths up to F9.8 = 3. Here we focus on the range up to F9.8 = 2, which is the main region where our models sit. In \fg{obser_10micron}, our fiducial model (blue dot) falls significantly below the observations. Varying model parameters (discussed in \sect{parameter}) does not improve the agreement with observations (empty circles). Thus, our base setup for the fiducial model may not agree with the nature of the inner disk. We note that the values for F9.8 agree reasonably well with observations, but the F11.3/F9.8 values are off. It means that the crystalline silicate feature at 11.3 $\mu$m is too strong with respect to the overall strength of the silicate feature. In the following section, we suggest an extended parameter study of models to improve the agreement with observations. 

\subsection{The nature of the innermost disk grains.}
\label{sec:ext_param}
\paragraph{}
Reducing the crystallinity in the innermost 0.1-0.2 au is a possible solution for the mismatch in the 10 $\mu$m shape. In \fg{testModel}, the crystallinity is the same everywhere in a disk with 0 \% and 10 \%, significantly lower than the inner disk of the fiducial model. 10 \% crystallinity (dashed line) shows much weaker forsterite features than the fiducial model (dotted line), and 0 \% crystallinity (solid line) shows a fully amorphous spectrum. The measurements for the  10 $\mu$m shape match observations well (\fg{obser_10micron}). Since the 10 $\mu$m feature is dominated by small inner disk dust, the most obvious solution is to remove small crystalline silicate dust from the inner disk. This can be done in two ways: (1) lower the inner disk crystallinity, (2) remove small dust irrespective of its crystallinity. 

We introduce a new set of models to investigate these possibilities. Model(a) deals with lowering the crystallinity, and Model(b) and (c) deal with decreasing the abundance of small dust in the inner disk. These models are listed in Table \ref{table:10micron_param}. Model(a) has reduced crystallinity by lowering the crystallinity fraction of the crystal reservoirs to 50 \% and 10 \% compared to the fiducial model. Model(a) with 50 \% still has strong forsterite features while the model with 10 \% has much weaker forsterite features except for the 33 $\mu$m feature (\fg{10specmodels}a). Thus, the ratio between 9.8 $\mu$m and 11.3 $\mu$m is much greater for Model(a) with 10 \% with respect to 9.8 $\mu$m strength, and this model is better in agreement with observations than the model with 50 \% (\fg{obser_10micron}). Note that the 16 $\mu$m feature in Model(a) with 10 \% crystal reservoir disappears while the feature in \fg{testModel} with globally 10 \% crystallinity is still visible. This is a clue that the 16 $\mu$m feature is still sensing the inner disk.
\begin{figure}
    \centering
    \includegraphics[width=\linewidth]{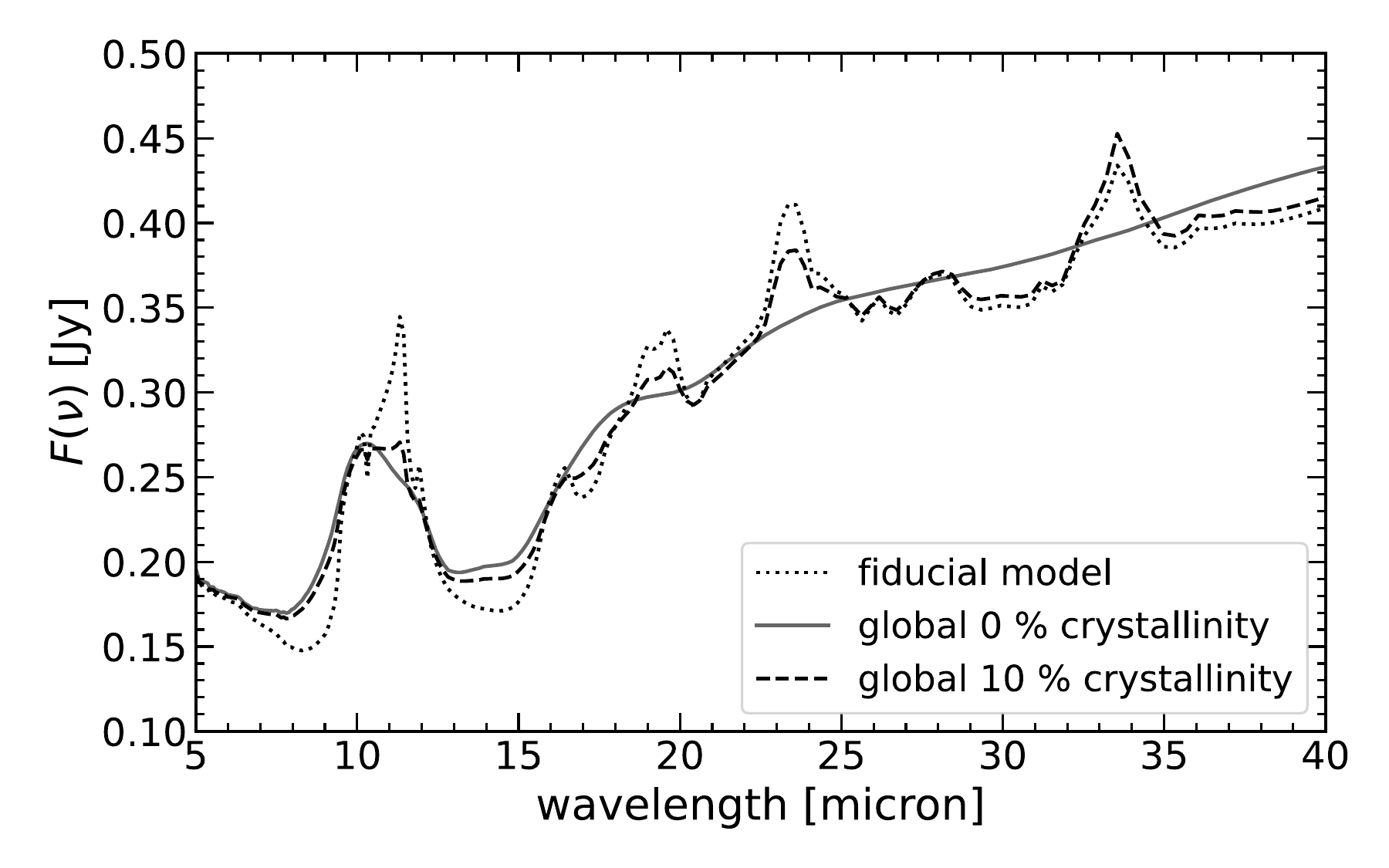}
    \caption{Model spectra for reduced disk crystallinity. Models have constant 0 \% (solid line) and 10 \% (dashed line) crystallinity all over the disk. }
    \label{fig:testModel}
\end{figure}

\begin{table}[]
\caption{Models to change the apparent crystallinity of the 10 $\mu$m silicate feature.}
\label{table:10micron_param} 
\centering     
\begin{tabular}{ccc}
\hline\hline
Model & parameter & variation \\ \hline
\multirow{2}{*}{(a)} & \multirow{2}{*}{\begin{tabular}[c]{@{}c@{}}crystallinity\\ of crystal reservoir\end{tabular}} & 50 \% \\
 &  & 10 \% \\
\multirow{2}{*}{(b)} & \multirow{2}{*}{sub-$\mu$m dust} & amorphous carbon \\
 &  & complete depletion \\
\multirow{2}{*}{(c)} & \multirow{2}{*}{$a_{\rm min}$} & 1 $\mu$m \\
 &  & 3 $\mu$m \\ \hline
\end{tabular}
\end{table}

\begin{figure*}
    \centering
    \includegraphics[width=\linewidth]{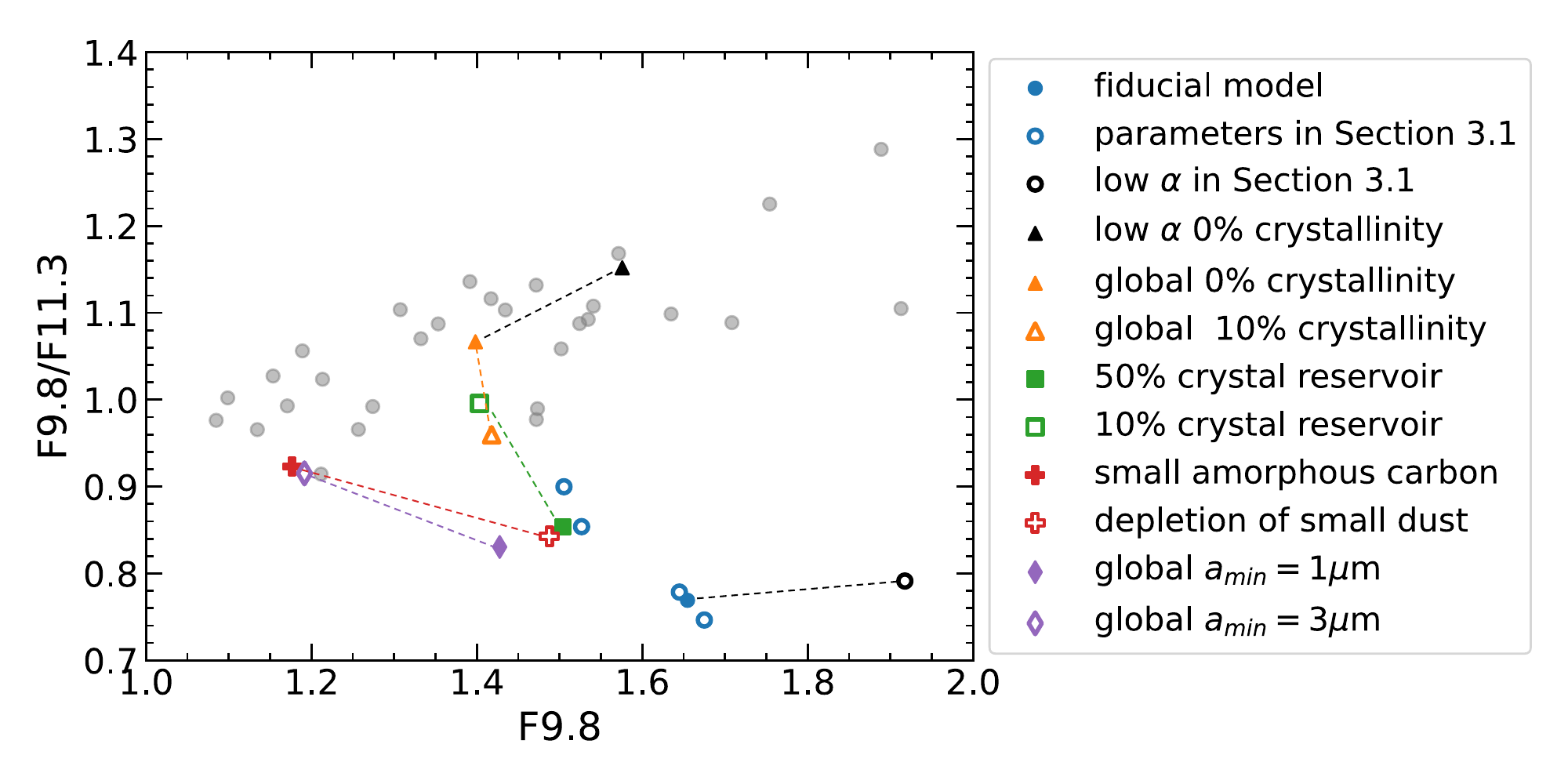}
    \caption{Correlation between the shape of 10 $\mu$m features (F9.8/F11.3) and the strength of the amorphous silicate feature at 9.8 $\mu$m (F9.8). Gray dots represent Spitzer IRS observations. The fiducial model is marked with a blue dot, and blue circles are the parameter space in \sect{parameter}. The models are summarized in Table \ref{table:10micron_param} and discussed in \sect{ext_param}.}
    \label{fig:obser_10micron}
\end{figure*}
 
\begin{figure}
    \centering
    \includegraphics[width=\linewidth]{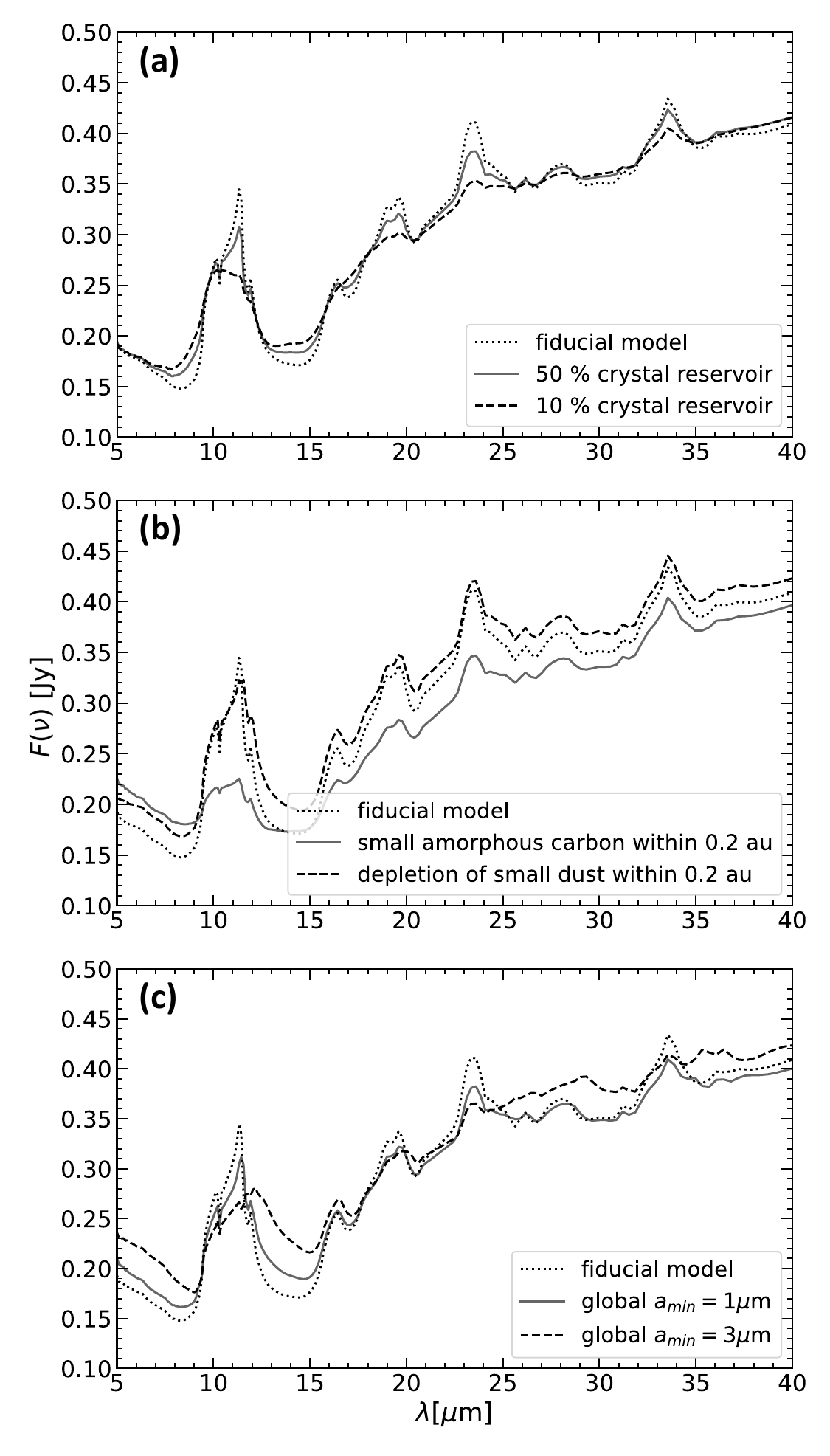}
    \caption{Model spectra for parameter study to reduce crystalline features in 10 $\mu$m silicate feature. Model(a) has crystal reservoirs with 50 \% (solid line) and 10 \% (dashed line) crystallinity instead of 100 \%. Model(b) is depleted in small silicate dust in the inner disk. The solid line shows a model that small silicate dust is replaced with featureless amorphous carbon, and the dashed line is a model that small dust in the inner disk is completely removed. Model(c) is depleted of small dust less than 1 $\mu$m (solid line) and 3 $\mu$m (dashed line) all over the disk.}
    \label{fig:10specmodels}
\end{figure}

For the depletion of small dust, irrespective of their lattice structure, we have two models. In Model(b), the small dust is only depleted in the inner disk ($r<0.2$ au) while small dust is depleted everywhere in the disk in Model(c). For Model(b), we replaced sub-micron ($a_{\rm min} < a < 1 \mu$m) dust with a featureless component to reduce the abundance of small silicate dust in the inner disk or completely depleted the sub-micron dust in the inner disk. In Model(b) with the featureless component, we used amorphous carbon as a proxy for the featureless component, so the opacity of the disk surface is the same as the fiducial model with a reduced abundance of sub-micron silicate dust. Because of the overall decrease in abundance of silicate, the 10 $\mu$m silicate feature is suppressed (\fg{10specmodels}b). Since the abundance of forsterite was significantly reduced, the 11.3 $\mu$m strength gets much weaker than the 9.8 $\mu$m strength (solid line). Thus, the measured 10 $\mu$m feature shape remains on the lower left compared to observed values in \fg{obser_10micron}. For Model(b) with complete depletion of small dust, we modified the abundance of small dust in the inner disk by removing sub-micron dust in the first five zones ($r<0.2$ au) in MCMax. It causes some discrepancy between the total dust mass and the dust mass of the inner zones, but the discrepancy is less than 2\% per zone. The discrepancy in dust mass is minor, but the opacity significantly changes. As a result, the 10 $\mu$m feature is weaker than the fiducial model but still strong (\fg{10specmodels}b). Hence, the shape still does not agree with the observations in \fg{obser_10micron}.

Further, we globally depleted small dust < 1 $\mu$m or 3 $\mu$m by setting $a_{\text{min}} =$ 1 $\mu$m or 3 $\mu$m in size distribution of dust in Model(c). In Model(c) with $a_{\text{min}} =$ 1 $\mu$m, a few micron dust still emits strong forsterite feature at 11.3 $\mu$m while Model(c) with $a_{\text{min}} =$ 3 $\mu$m shows much weaker 11.3 $\mu$m feature. This is because the forsterite features flatten with increasing the grain size (see Appendix \ref{sec:appendix}). Thus, Model(c) with $a_{\text{min}} =$ 3 $\mu$m agrees better with observations than Model(c) with $a_{\text{min}} =$ 1 $\mu$m in \fg{obser_10micron}. We have not explored variations of the slope of the grain size distribution ($a_{\rm pow}$). Qualitatively, we expect the effect of decreasing $a_{\rm pow}$ to be similar to the effect of removing the small dust grains from the disk surface, as shown, for instance, in Model(c), because the fraction of small grains decreases with respect to large grains.

In general, both reducing crystallinity and removing small dust result in 10 $\mu$m feature shapes that better match with observations. The crystallinity of the crystal reservoirs in the inner disk has to be significantly reduced from our fiducial model, as low as 10 \%. We find that reducing the abundance of small crystalline silicates in the inner disk ($<$ 0.2 au) is more effective in weakening the 11.3 $\mu$m crystalline silicate feature than just removing small dust in the inner disk. A significant depletion of small dust is needed in a disk with $a_{\text{min}} = 3 \mu$m.

In \fg{obser_10micron}, the low $\alpha$ model from \sect{parameter} has a larger F9.8 value than the fiducial model because the smallest dust grains dominate the disk surface due to enhanced settling and produce a strong silicate feature (black empty circle). When we assume 0 \% crystallinity in the low $\alpha$ model, it agrees well with the observations (black triangle) and shifts to the upper right with respect to the 0 \% crystallinity model (orange triangle). The strength and shape of the 10 $\mu$m silicate feature vary by changing the disk turbulence. However, very low disk turbulence ($\alpha = 10^{-4}$) is not able to reproduce the strongest feature strength that we see in the Spitzer IRS observations. Thus, removing small dust is still needed to reproduce observations with a low F9.8 value.

\section{Discussion}
\label{sec:discussion}
\paragraph{}
In this section, we first compare our results with previous studies on the spatial distribution of crystalline silicates by thermal annealing and radial transport processes. Observations of crystalline silicates are also compared with our models, and we propose possible mechanisms that could explain models in \sect{ext_param}. The limitation of our models are also discussed.

    \subsection{Comparison with earlier works}
    \label{sec:earlywork}
    \paragraph{}
    \cite{Gail2001} studied thermal annealing processes and the radial transport of crystalline silicates in a protoplanetary disk. The paper defined the probability distribution for the degree of crystallization based on the disk temperature and calculated the average crystallinity of forsterite as a function of distance taking into account radial transport. As a result, a dust grain is either fully crystalline or fully amorphous, which corresponds to our assumption, and the relative abundance of crystal grains gradually decreases with increasing radial distance with the fully crystalline region in the innermost disk. The overall shape of the spatial distribution is similar to our models in \fg{radial_dist_all}. However, the value for the crystallinity is generally higher than our models. The author showed $10 - 50 \%$ crystallinity in $1 - 10$ au while our models show the crystallinity at $r < 1$ au. This is because we have a much smaller size of the crystal reservoirs and larger grains in our model. In \cite{Gail2001}, the region for the fully crystalline disk is up to $\sim$ 0.3 au for a disk with the lowest accretion rates of $10^{-8}$ M$_{\odot}$ yr$^{-1}$. Most of our models have the outer edge of crystal reservoirs $r_{\rm cr} < 0.2$ au with the accretion rate of $\sim 10^{-9}$ M$_{\odot}$ yr$^{-1}$, and only the smallest dust grains in the low $\alpha$ model show slightly larger $r_{\rm cr}$. Moreover, the author follows the grain size distribution of \cite{Mathis_etal1977} with $a_{\rm l} = 0.005 \mu$m to $a_{\rm u} = 0.25 \mu$m, which is much smaller than our grain size distribution in Table \ref{table:properties}. If our model had only sub-micron grains, the fraction of crystallinity would become more comparable with \cite{Gail2001}. 
    
    \cite{Arakawa_etal2021} also investigated the radial transport of crystalline dust grains in a disk. They describe how different sizes of crystalline dust grains are distributed to the outer disk. Their analysis considers the gas pressure structure influenced by a disk wind. The gas pressure radially increases in the inner disk up to the pressure maximum and then decreases. Due to this pressure gradient, dust grains in the inner disk experience outward drift. Large dust grains, decoupled from the gas, efficiently drift outward, resulting in higher crystallinity in the outer disk for large grains than our results. In our study, the gas pressure only decreases with increasing $r$, so we do not consider the outward drift. Thus, small dust grains are distributed to the outer disk more efficiently than large dust grains in our models.

    \subsection{Confronting our crystallization model with observations}
    \label{sec:confrontingObs}
    \paragraph{}
    The region where the crystalline silicates exist and its contribution to the observed spectra have been investigated observationally in previous studies. \cite{Watson_etal2009} studied 84 T Tauri disks \citep{Furlan_etal2005,Furlan_etal2006}, observed by Spitzer. From the analysis of the full infrared spectrum (5.3-38 $\mu$m), the authors found that the appearance of crystalline silicate features implies the existence of crystalline silicates within the inner disk ($r \lesssim 10$ au). In our model, crystalline silicates from the crystal reservoirs are radially distributed up to several au after 1 Myr. In \fg{contribution_fidu}, features around $10 \mu$m are dominated by the inner disk $\lesssim 1$ au while the outer disk ($r>10$ au) contributes to features at $ \lambda > 20 \mu$m more than $20 \%$. \cite{Olofsson_etal2009} also studied more than a hundred T Tauri disks observed by Spitzer and found crystalline silicates in 96 disks. The authors argued the warm inner disk emits features at $\lambda \sim 10 \mu$m and the outer cold disk emits $\lambda > 20 \mu$m, which qualitatively corresponds to our results. Moreover, the fact that such a high fraction of disks shows crystalline silicates is also in agreement with our models in a wide range of parameters, that show detectable crystalline silicates.
    
    Disks with high crystallinity may refer a low disk turbulence. \cite{Olofsson_etal2009} suggested a weak anticorrelation between spectral slope ($F_{30}/F_{13}$: ratio of fluxes in the range 29-31 $\mu$m and in the range 13 $\mu$m) and crystalline feature strength around 23 $\mu$m. As $F_{30}/F_{13}$ is small, representing a blue SED slope, the crystalline feature gets stronger. The blue SED slope can be seen for a ﬂattened disk with low turbulence because the optically thin disk surface is warmer, causing a larger $F_{13}$ value \citep{Olofsson_etal2009}. This result agrees with our low disk turbulence model, showing high crystallinity with stronger feature fluxes in general.

    \subsection{Scenarios to reduce 10 $\mu$m forsterite feature in the model}
    \label{sec:senarios}
    \paragraph{}
    We note that models with reduced crystallinity everywhere in a disk or only in the inner disk show 10 $\mu$m feature shapes that qualitatively match the observed shapes. One possible mechanism to reduce the inner disk crystallinity is amorphization by stellar radiation in the disk surface \citep{Glauser_etal2009}. X-ray emission from the central star breaks the lattice structure of crystalline silicates, so the dust becomes amorphous. This amorphization is more effective for dust in the optically thin disk surface and the inner disk. The rate of amorphization compared to crystallization has to be further studied to determine how effective this mechanism is to reduce inner disk crystallinity. 
    
    The inner disk could be depleted in small silicate species or overall small dust. Due to the high temperature of the innermost disk, small silicate dust may not survive because these grains sublimate. \cite{Kama_etal2009} showed effective sublimation of small grains, which changes the location of the optically thin inner rim in a disk around a Herbig Ae star. In their model that only contains 0.1 $\mu$m grains, the optically thin inner rim is beyond 2 au while it is within 0.5 au for 10 $\mu$m and 100 $\mu$m grains. Another mechanism to deplete a disk of small grains are disk winds. \cite{Miyake_etal2016} explore the radial distribution of grain sizes within a disk experiencing magnetorotational turbulence-driven disk wind. They argue that 1 micron dust is expelled by the vertical gas outflow in the inner disk $r \lesssim 10$ au. Interestingly, a dusty disk wind dominated by small grains was recently inferred from scattered light observations of the disk of RY~Tau \citep{Valegard_etal2022}
    
    Instead, other refractory species, such as metallic iron, could still remain in the silicate-depleted region. As Model(b) with amorphous carbon shows, the presence of dust species other than silicate in the inner disk could affect the shape of the 10 $\mu$m feature. In addition, the dust evolves to a large size by coagulation and could cause depletion of small dust globally. \cite{Dullemond_Dominik2004b} modeled dust coagulation in protoplanetary disks. In their simple one-particle models, sub-$\mu$m dust grows to a few $\mu$m dust at 1 au within 800 years. Model(c) with the larger $a_{\text{min}}$ could explain the growth of smaller dust to larger dust. Moreover, \cite{Birnstiel_etal2012} simulated the global evolution of dust in protoplanetary disks and showed the evolution of the surface density as a function of grain size (see their Fig 1 and Fig 2). Smaller dust gets depleted as the disk evolves up to a fragmentation regime in the inner disk and radial drift regime in the outer disk within a few Myr. \cite{Dominik_Dullemond2024} applied the bouncing barrier to dust evolution and explored the size distribution of dust within their modeled protoplanetary disk. Their investigation revealed a highly efficient depletion of micron-sized dust and the limitation of the largest grain size, resulting in a quasi-monodisperse size distribution. In their turbulent disk with $\alpha = 10^{-3}$, the typical grain size at 1 au was found to be in the range of tens of microns. Extrapolating to our turbulent disk with $\alpha = 10^{-2}$, we anticipate smaller grain sizes. Moreover, \cite{Arakawa_etal2023} discuss the efficiency of collisional growth, considering both bouncing and sticking mechanisms. The findings indicate that smaller grains are more likely to grow through collisions, potentially leading to the depletion of small grains.

    For simplicity, we only considered forsterite for the crystalline silicates in this paper. In protoplanetary disks, several types of crystalline silicates have been observed, such as enstatite and silica \citep{Bouwman_etal2001, Watson_etal2009, Olofsson_etal2009}. These different species have different activation energies to be crystallized, so the region for the crystallization region changes and so does the crystal reservoir. The crystal reservoir is one of the factors to determine the radial distribution of crystalline silicates. Thus, different species would have their own spatial distributions, which would cause different mid-infrared spectra. \cite{Bouwman_etal2008} found different relative abundances of enstatite (MgSiO$_3$) and forsterite in the inner and outer disks of seven protoplanetary disks, observed by Spitzer. A similar result was found for a larger sample of disks around more massive Herbig Ae/Be stars \citep{Juhasz_etal2010}. Within 1 au, enstatite is more abundant than forsterite and vice versa within 5-15 au. The observed spectra would then represent the distributions of crystalline silicate species. Further study with a set of different dust species would be needed to explain the different radial abundances of observed dust species in a disk.

\section{Conclusion}
\label{sec:conclusion}
\paragraph{}
In this paper, we modeled the spatial distribution of crystalline silicates in planet forming disks. We first defined the region where the amorphous silicates get crystallized (crystallization region). From the normalized scale height, the Stokes number of gas-coupled dust was estimated in order to determine how different grain sizes can be vertically mixed (dust layer). The region where the dust reaches the crystallization region is defined as a crystal reservoir ($r_{\rm cr}$). We applied DISKLAB to constrain the radial distribution of crystalline silicates in the midplane. Based on the distribution, we modeled the mid-infrared spectrum using the radiative transfer code MCMax. The main conclusions are listed below: 

\begin{itemize}
    \item Observed small dust in the disk surface represents the distribution in the midplane. The vertical mixing timescale is much shorter than the radial transport timescale within a few au for $\alpha=10^{-2}$ and $10^{-3}$ disks. Thus, the dust gets vertically well mixed before the radial distribution changes in the midplane. 
    \item The inner disk contributes more to features at shorter wavelengths; the outer disk contributes more to features at longer wavelengths. 10 $\mu$m forsterite feature is dominated by the inner disk $< 0.2$ au while 33 $\mu$m feature is dominated by the outer disk $> 0.2$ au (\fg{contribution_fidu}). 
    \item When the initial crystallinity by any early evolution of the disk changes, the overall strengths of features change together. Long wavelength features are more sensitive to this because the outer disk is the main region where the initial crystallinity changes.  
    \item With higher Sc, meaning weaker diffusion of dust, the overall band strengths become weaker because the disk beyond the crystal reservoir has a lower crystallinity than the fiducial model due to lower dust diffusivity. 
    \item A change in the initial crystallinity affects longer wavelengths more, while Sc affects shorter wavelength bands. Thus, increasing initial crystallinity together with Sc results in weaker short-wavelength features and stronger longer-wavelength features than the fiducial model. 
    \item Lower disk turbulence still provides strong features despite of smaller $r_{\rm cr}$. Because dust gets easily decoupled from the gas in a quiescent disk, its disk surface is dominated by the smallest dust grains, which produce strong dust features.
    \item Modeled crystalline features with forsterite qualitatively agree with observed spectra except for the 10 $\mu$m silicate feature because our standard model overpredicts the crystallinity in the innermost disk. 
    \item We find 100 \% crystallinity of the crystal reservoir and a gradual decrease of crystallinity beyond $r_{\rm cr}$ is too simplistic and does not agree with the observations. We need a reduced abundance of small crystalline silicate in the innermost disk (less than 0.2 au) to match observations. Possible mechanisms are amorphization, sublimation of silicate, and growth of dust.
\end{itemize}

Further study should address the effect of amorphization on the crystallinity of the crystal reservoirs by comparing the rate of amorphization and crystallization. The inner disk structures are not well understood yet, so the sublimation of silicate dust and spatially changing grain size distribution coming out of dust evolution in the inner disk should also follow for further study. Moreover, including various species in the dust-not just forsterite-would provide a better understanding of the mid-infrared observation in further work. 

\section*{Acknowledgments}
We appreciate the anonymous referee for their insightful comments. We acknowledge to Combined Atlas of Sources with Spitzer IRS Spectra (CASSIS) database for the use of low-resolution spectra. H.J. thanks M. Min for the use of MCMax radiative transfer code and C. Dominik for the fruitful discussion. 

\bibliographystyle{aa}
\bibliography{ref}

\appendix
\section{Opacities of different grain sizes}
\label{sec:appendix}
\paragraph{}
Dust opacities tend to decrease as dust grain sizes increase due to the increase in the volume to surface ratio, resulting in  weaker peaks of the resonances for larger grains, as illustrated in \fg{comp_opa}. The absorption opacities of 1 $\mu$m and 3 $\mu$m forsterite grains are obtained using the Distribution of Hollow Sphere model \citep{Min_etal2005}. The 3 $\mu$m grain exhibits weaker features compared to the 1 $\mu$m grain. Consequently, depleting small dust particles from the disk surface results in weaker crystalline features in the modeled spectra in \sect{ext_param}.
\begin{figure}[h]
    \centering
    \includegraphics[width=\linewidth]{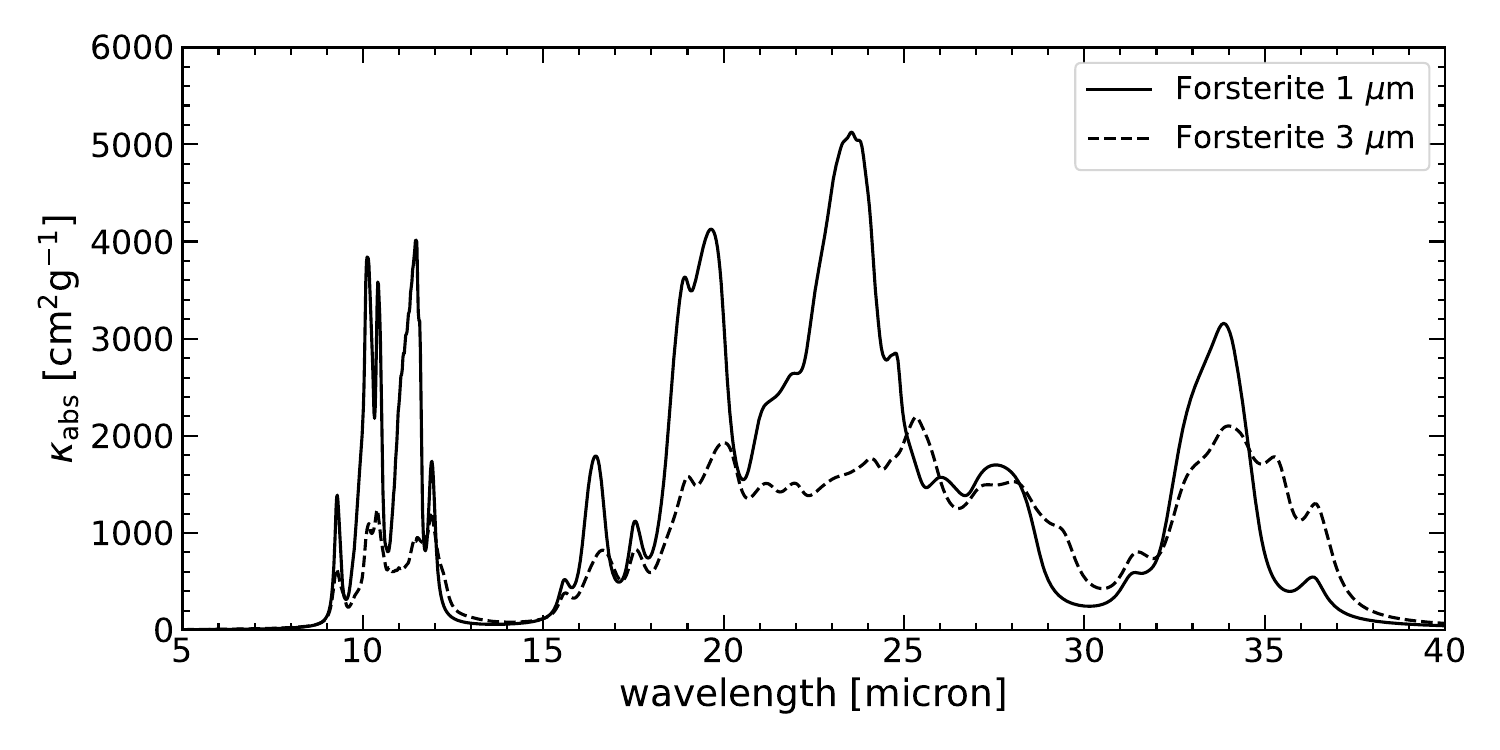}
    \caption{Absorption coefficients of 1 $\mu$m (solid line) and 3 $\mu$m (dashed line) forsterite grains. The 1 $\mu$m grain exhibits more prominent dust features compared to the 3 $\mu$m grain.} 
    \label{fig:comp_opa}
\end{figure}

\end{document}